\documentclass[conference]{IEEEtran}
\IEEEoverridecommandlockouts
\newcommand{\eg}{{\it e.g., }}
\newcommand{\ie}{{\it i.e., }}
\newcommand{\etal}{{\it et~al. }}
\usepackage{url}
\usepackage{cite}
\usepackage{comment}
\usepackage{amsmath,amssymb,amsfonts}
\usepackage{algorithmic}
\usepackage{graphicx, caption}
\usepackage{textcomp}
\usepackage{balance}
\usepackage{soul}
\usepackage{stfloats}
\usepackage{multirow}
\usepackage{amsmath, amssymb}
\usepackage{subfig}
\usepackage{xcolor}
\usepackage[colorlinks, citecolor=blue]{hyperref}
\usepackage{cleveref}
\usepackage[utf8]{inputenc}
\usepackage{enumitem}
\crefformat{subsection}{\S#2#1#3}

\newlength{\boxfigwidth}
\newcommand{\boxfig}[1]{
\begin{figure}[ht!]
\begin{center}
\setlength{\boxfigwidth}{0.42\textwidth}
\addtolength{\boxfigwidth}{0in}
\noindent\framebox{\colorbox{blue!10}{\quad
\begin{minipage}{\boxfigwidth}
#1
\end{minipage}\quad}}
\end{center}
\end{figure}
}

\begin{document}

\title{Exploring the Impact of Virtualization on the Usability of the Deep Learning Applications \\

\thanks{This research is supported by the National Science Foundation under award\# CNS-2007209. Special thanks to our colleagues SM Zobaed and Connor M. Rawls for their help in conducting this study.}
}

\author{

\IEEEauthorblockN{Davood G. Samani, Mohsen Amini Salehi}
\IEEEauthorblockA{\textit{High Performance Cloud Computing (HPCC) Lab,}
\textit{University of Louisiana at Lafayette, USA}\\
\{d.ghatrehsamani1,amini\}@louisiana.edu}


}

\maketitle

\begin{abstract}

Deep Learning-based (DL) applications are becoming increasingly popular and advancing at an unprecedented pace. While many research works are being undertaken to enhance Deep Neural Networks (DNN)---the centerpiece of DL applications---practical deployment challenges of these applications in the Cloud and Edge systems, and their impact on the usability of the applications have not been sufficiently investigated. In particular, the impact of deploying different virtualization platforms, offered by the Cloud and Edge, on the usability of DL applications (in terms of the End-to-End (E2E) inference time) has remained an open question. Importantly, resource elasticity (by means of scale-up), CPU pinning, and processor type (CPU vs GPU) configurations have shown to be influential on the virtualization overhead. Accordingly, the goal of this research is to study the impact of these potentially decisive deployment options on the E2E performance, thus, usability of the DL applications. 
To that end, we measure the impact of four popular execution platforms (namely, bare-metal, virtual machine (VM), container, and container in VM) on the E2E inference time of four types of DL applications, upon changing processor configuration (scale-up, CPU pinning) and processor types. This study reveals a set of interesting and sometimes counter-intuitive findings that can be used as best practices by Cloud solution architects to efficiently deploy DL applications in various systems. The notable finding is that the solution architects must be aware of the DL application characteristics, particularly, their pre- and post-processing requirements, to be able to optimally choose and configure an execution platform, determine the use of GPU, and decide the efficient scale-up range.

\end{abstract}


\section{Introduction}
Deep Learning (DL) applications, operating based on Deep Neural Networks (DNN), are increasingly being adopted in the IT world. Companies, such as  Google, Facebook, Microsoft, Baidu, and many others are aggressively leveraging DL to instill intelligence into their products.

At the production level, just like any other application, DL applications are deployed in isolated environments (known as \emph{execution platforms}) that are offered by the Cloud and Edge 
computing platforms. In its simplest form, an execution platform can be a bare-metal (BM) server. However, for cross-application isolation, often, more complicated execution platforms in form of conventional Virtual Machine (VMs) \cite{salehi2014resource}, lightweight Micro-VM (MVM) \cite{agache2020firecracker}, container (CN) \cite{pahl2015containerization}, or container on VM (VMCN) \cite{samani2020art}, are offered by the Cloud providers. Each execution platform is composed of different abstraction layers, and each layer is expected to impose an overhead that inevitably causes some level of performance degradation for the application. 

\begin{figure}[ht]
  \centering
  \vspace{-0.1in}
  \includegraphics[width=0.45\textwidth]{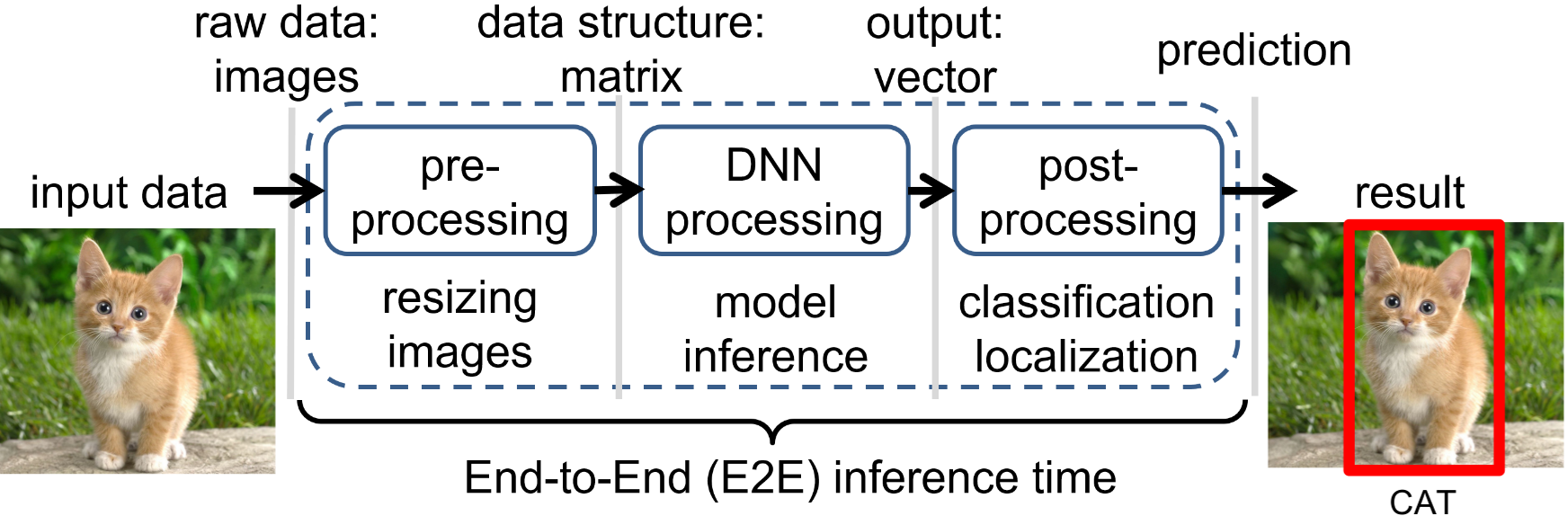}
  \caption{\small{Three operations involved in an inference request. The type of data transferred between each stage is shown at the top, and an example of image classification application is provided at the bottom. }}
  \label{Fig:inferenceStages}
   \vspace{-0.05in}
\end{figure}

The imposed overhead is particularly crucial for the \emph{inference} operation of DL applications. In fact, the \emph{training} operation is generally an offline process, conducted on highly parallel (\eg GPU) Cloud servers. In contrast, the inference operation is often an online and latency-sensitive process~\cite{lazaro2019approach} performed in the wild, on a wide range of systems---from powerful Cloud servers to resource-constrained Edge devices. As shown in Figure~\ref{Fig:inferenceStages}, each inference request in a DL application entails three operations: \emph{pre-processing} input, performing the \emph{inference} operation, and \emph{post-processing} the result. For instance, in an image classification application, pre-processing is to resize the input image based on the dimensions dictated by the DNN model; The inference is to run the DNN model; And, post-processing is to match the inference result to a certain object in the classification. Accordingly, for a given inference request, the \emph{End-to-End (E2E) inference time} is the total time to pre-process input, perform inference, and post-process the result \cite{devlin2018bert}. 

Even a minor difference in the imposed overhead of execution platforms can be decisive on the E2E inference time and can affect the \emph{usability} of latency-sensitive DL applications in the production environment. This imposed overhead raises a fundamental question for Cloud solution architects on \emph{what is the optimal execution platform to deploy a DL application?} This question gets further complicated when we know that the \emph{configurations of each execution platform can significantly impact its overhead}. In particular, it is proven that the inference operation is extensively based on processor-intensive matrix calculations \cite{reddi2020mlperf}, hence, we are to explore the processor-related configurations of the execution platforms. Moreover, characteristics of the DL applications can influence the overhead of the underlying execution platform. Thus, we consider several DL applications that are different in terms of: (A) the type of DNN model utilized at their kernel; and (B) the type of their input data (\eg image vs text).

To design an effective deployment plan for a given DL application, a solution architect should know the answer to the following three configuration questions:
\begin{enumerate}
    \item How does the scale-up level (\ie the number of processing cores) of an execution platform impact its overhead?
    \item One way to potentially mitigate the overhead of an execution platform is to configure it with CPU pinning \cite{PinningHPCCLab}. How much such a configuration can mitigate the overhead of an execution platform?
    \item Last but not least is to configure an execution platform on heterogeneous machines, particularly, the popular CPU-GPU combination. How can such a configuration impact the imposed overhead of an execution platform? 
\end{enumerate}

To answer these questions, we conduct a comprehensive analysis via examining various execution platforms, offered by Cloud and Edge, to run a diverse set of DL applications with different DNN models, and different input types. For that purpose, we utilize MLPerf \cite{reddi2020mlperf}, a widely-known benchmarking tool for DNN-based applications, that includes: Image Classification \cite{reddi2020mlperfVision}, Speech Recognition \cite{reddi2020mlperfSpeech}, Natural Language Processing (NLP) \cite{reddi2020mlperfLanguage}, and Recommendation Systems \cite{reddi2020mlperfRecommendation}. We define a \emph{scenario} as a DL application deployed within a certain execution platform with a specific configuration. We use the E2E inference time within each execution platform as the metric to measure the imposed overhead. 
\begin{figure*}[t]
  \centering
  \vspace{-0.2in}
  \includegraphics[width=0.77\textwidth]{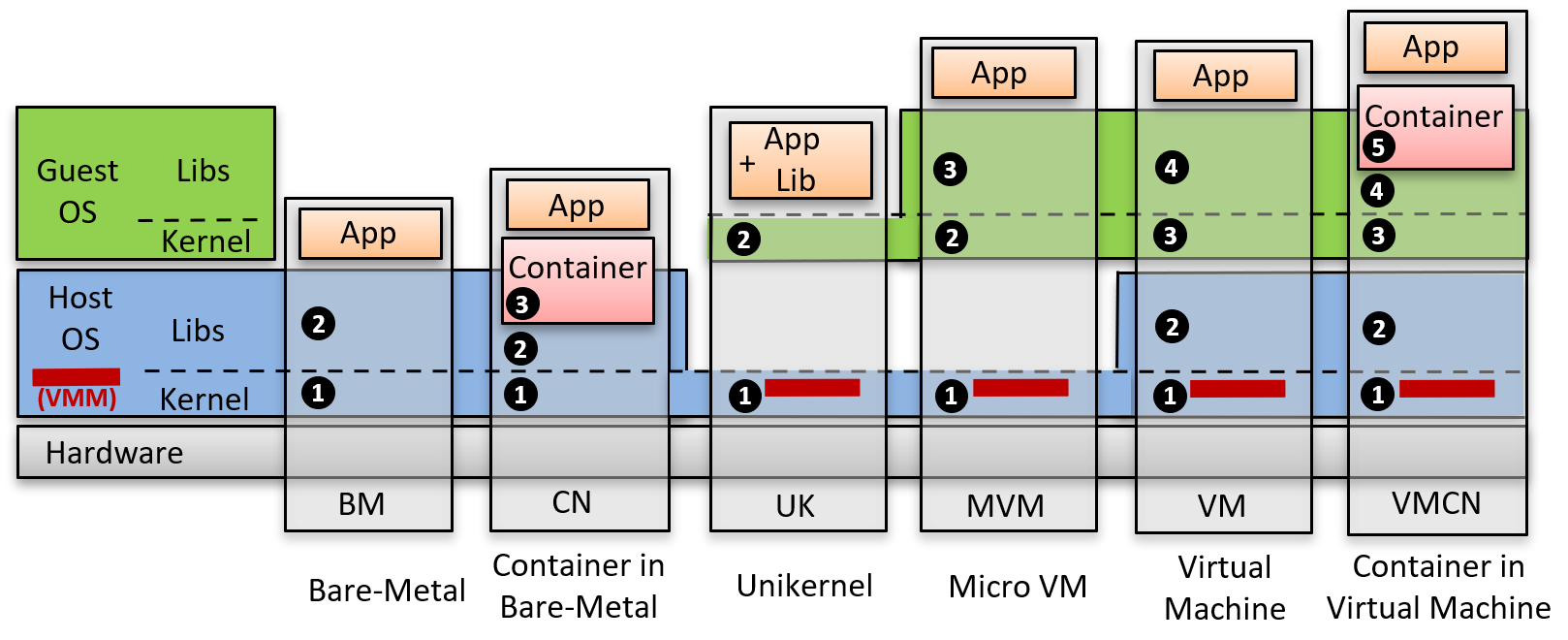}
  \caption{\small{Various execution platforms offered by Cloud providers and their constituent abstraction layers. Each circled number represents one abstraction layer in a platform. The thickness of each layer (horizontal ribbons) implies the magnitude of overhead imposed by that layer.}}
  \label{Fig:Platforms}
   \vspace{-0.2in}
\end{figure*}
In summary, the contributions of this work are as follows:
\begin{itemize}
    \item Quantifying and analyzing the performance overhead imposed by different execution platforms to perform E2E inference in DL applications.
    \item Provided the processor-intensive nature of the inference operation, we study the influence of processor configurations of each execution platform, namely processor scale-up, processor pinning, and processor types (CPU vs GPU), on the imposed overhead.
    \item Analyzing the influence of DL applications' characteristics on the execution platform overhead.
    \item Providing a set of best practices for solution architects to deploy DL applications in a production environment.
\end{itemize}

The rest of this paper is structured as follows: Section \cref{sec:background} provides a background of DL concepts and virtualization technologies followed by a review of the related works. Then, Section \cref{sec:evaluation} describes the specification of DL applications used in the benchmarking, execution platforms, and evaluation scenarios. Section \cref{sec:applicationSpecific} explains our application-specific analysis and insights. A cross-application analysis focusing on the repetitive patterns is discussed in Section \cref{sec:crossAnalysis}. Lastly, Section \cref{sec:summary} summarizes our findings and lists a set of best practices for efficient deployment of execution platforms for DL applications.

\section{Background and Related Studies}
\label{sec:background}
Virtualization platforms (\ie execution platforms minus bare-metal) abstract the host resources, such as compute, storage, and network, to enable isolated sharing of the virtual resources across multiple users. As depicted in Figure \ref{Fig:Platforms}, current virtualization technologies are categorized based on the level of abstraction they offer. Hardware virtualization (a.k.a. VM) enables multitenancy  on a physical host via sharing virtual resources across multiple isolated guest operating systems. It relies on a \textit{hypervisor} (a.k.a. \emph{VMM}) to abstract physical resources (\eg CPU) using the kernel modules and offer virtual resources (\eg vCPU). It also uses an emulator (\eg \texttt{virtio} in KVM) at the host OS level to facilitate virtual IO for VMs. More recently, lightweight hypervisors, known as Micro VMs (MVM), that eliminate the emulator layer were developed. Importantly, AWS FireCracker \cite{agache2020firecracker} is a MVM that enables VM-level isolation for AWS Lambda functions \cite{denninnart2021harnessing}. Unikernel (UK) \cite{madhavapeddy2013unikernels} is another VM-based technology that, in addition to a lightweight hypervisor, takes advantage of a lightweight guest OS that only includes application libraries. Although Unikernel appears to be a low-overhead platform, due to the implementation complexity, it has yet to be utilized for DL applications \cite{leon2020dark}.

Containerization \cite{denninnart2021harnessing} is an OS-level virtualization technique where the host OS kernel is shared across multiple isolated user-space environments, called \textit{namespaces}. Unlike VMs, there is no resource abstraction layer for namespaces that makes containers lightweight. As shown in Figure \ref{Fig:Platforms}, containers (henceforth termed, \emph{CN}) can be either deployed directly on a bare-metal (\emph{BM}) server or within a VM (\emph{VMCN}).

In both hardware- and OS-level virtualizations, time-sharing scheduling policies, such as Completely Fair Scheduler (CFS), are used at the host OS level. Therefore, the CPU cores allocated to a given VM/container can vary after each scheduling quantum. In this study, we call this CPU provisioning fashion as the \textit{vanilla} mode. Although the overhead of vanilla CPU provisioning is negligible in many cases, it can become a challenge for latency-sensitive applications running in hosts with multiple CPU sockets. To overcome this challenge, Cloud architects commonly override the default host OS scheduler and allocate certain CPU cores to a given VM/container, which is known as \emph{CPU Pinning} \cite{PinningHPCCLab}. We note that, unlike vanilla mode that utilizes all the host CPU cores to cumulatively offer the expected performance, in the pinned mode, only the designated cores are utilized and the rest are left idle.

Several research works \cite{potdar2020performance,lingayat2018performance,samani2020art} have been undertaken to evaluate the overhead of different execution platforms. Specifically, our prior study \cite{samani2020art} investigates various application types (MPI, big data, and FFMpeg) across different execution platforms. However, these prior works neither consider DL inference as their workload, nor do they study the impact of scale-up and heterogeneity on the E2E inference time.
Several other research works have been conducted to benchmark: (A) the training and inference time of various  DNN models on various DL back-end frameworks \cite{devarajan2021dlio,koirala2019deep,duan2016benchmarking,mase2020benchmarking,zobaed2021saed,fagbohungbe2021benchmarking,liu2018benchmarking}; and (B) the underlying hardware platforms, both at the system-level (\eg Cloud, Edge, and device) \cite{dai2019benchmarking}, and at the processor-level (\eg CPU, GPU, TPU, and FPGA) \cite{wang2019benchmarking}. Nonetheless, only a few works have centered their research on the impact of virtualization platforms. 
The most related study was conducted by Lin \etal \cite{lin2018comparison} where they investigated the impact of BM, CN, and VM execution platforms on the training of a DL application for image classification using different DNN models. In contrast, our study concentrates on the inference operations of various applications types under different configurations and scenarios depicted in Figure \ref{Fig:scenarios}.

\begin{table*}[b]
\begin{center}
\begin{tabular}
{c|c|c|c|c}
\hline
  \textbf{DL Application} &
  \textbf{Image Classification (IC)} &
  \textbf{Speech Recognition (SR)} &
  \textbf{NLP} &
  \textbf{Recomm. Systems (RS)} \\ \hline \hline
\textbf{Input Data Type}   & Image                      & Audio               & Text                          & Large volume text \\ \hline
\textbf{NN Model}          & ResNet50-v1.5              & RNNT               & BERT Large                    & DLRM (Debugging)  \\ \hline
\textbf{No. of Parameters}        & 25.6 Millions              & 45.3 Millions       & 340 Millions                  & Not mentioned     \\ \hline
\textbf{Precision}         & fp32                       & fp32                & fp32                          & fp32              \\ \hline
\textbf{Model Size}        & 97.5MB                     & 519MB               & 4GB                           & 1GB               \\ \hline
\textbf{Back-end Framework} & TensorFlow                 & PyTorch             & TensorFlow                    & PyTorch           \\ \hline
\textbf{No. of Inference Req.}           & 1000                       & 1200                & N/A                           & N/A               \\ \hline
\textbf{Test Dataset}         & ImageNet2012               & OpenSLR (dev-clean) & Validation Dataset SQuAD v1.1 & Criteo Kaggle DAC \\ \hline
\textbf{Evaluation Context} & Cloud: CPU,GPU - Edge: CPU & Cloud: CPU          & Cloud: CPU, GPU                    & Cloud: CPU, GPU*  \\ \hline
\end{tabular}
\vspace{0.1in}
\caption{\small{Characteristics of the Deep Learning (DL) applications offered by MLPerf. For IC and SR, MLPerf allows us to limit the number of generated inference requests, whereas, for NLP and RS, MLPerf uses the entirety of its test dataset, hence number of inference requests for these two applications is N/A. The asterisk (*) indicates that the container (CN) execution platform is not supported for the RS application.}}
\label{table:AppCharacteristics}
\end{center} 
\end{table*}

\section{Evaluation Methodology and Experimental Setup}
\label{sec:evaluation}
\subsection{Deep Learning Benchmarking Tools}\label{AA}

Examining different execution platforms to deploy DL applications requires standard benchmarking tools that have popular DL applications as their workload and can collect their performance information. Several benchmarking tools, such as \texttt{MLPerf} \cite{reddi2019mlperf}, \texttt{DLBS} \cite{dlbsGithub}, and \texttt{DeepBench} \cite{deepbenchGithub}, are available that each one focuses on a different type of workload and measures certain performance metrics. In particular, MLPerf is widely-used benchmarking tool that offers a collection of six DL applications with different DNN models and various input types. 
Deep Learning Benchmarking Suite (DLBS) \cite{dlbsGithub} is a benchmarking tool that focuses on measuring the inference performance using different back-end frameworks (\eg TensorFlow and PyTorch) for DL applications. Unlike MLPerf that measures the E2E inference performance, DLBS captures only the inference time and disregards the pre- and post-processing steps, included in a DL application deployment. Accordingly, in this study, we primarily use MLPerf to measure the E2E inference time of DL applications and occasionally use DLBS to concentrate on the impact of the DNN models\footnote{DeepBench \cite{liu2018benchmarking} is another benchmarking tool that measures the performance of underlying OS by capturing the most time-consuming DL operations in the kernel and user spaces, which is out of the scope of this study.}.

As detailed in Table~ \ref{table:AppCharacteristics}, we configure MLPerf to measure the E2E inference time of the following four DL applications: 
\begin{enumerate}
    \item \emph{Image Classification (IC)}: For a given input image in the \texttt{jpg} format, the IC application determines the probabilities that it belongs to different classes of objects. It uses the ResNet50 DNN model \cite{reddi2020mlperfVision} of Tensorflow. The MLPerf benchmaking workload includes 1,000 inference requests.
    
    \item \emph{Speech Recognition (SR)}: It detects the character transcriptions of an input sound file with the \texttt{wav} format. It operates based on the PyTorch RNN-Transducer (RNNT) \cite{reddi2020mlperfSpeech} model. The workload includes 1,200 15-second-long audio files from OpenSLR (dev-clean) dataset. 
    
    \item \emph{Natural Language Processing (NLP)}: NLP applications are to analyze a written text and understand the meaning of it. MLPerf uses the BERT DNN model \cite{reddi2020mlperfLanguage} as its NLP application, and SQuAD v1.1 is used as its test dataset. 

    \item \emph{Recommendation Systems (RS)}: Recommendation systems predict users preferences based on their prior interactions. MLPerf uses a PyTorch-based model, known as Deep Learning Recommendation Model (DLRM) \cite{reddi2020mlperfRecommendation}, as its RS application. It employs the Criteo Kaggle Display Advertising Challenge (DAC) as its dataset. 

\end{enumerate}  

To make the imposed overhead of each execution platform visible, we put them under stress load via feeding a batch of inference requests (a.k.a. workload) to each DL application and report the time to complete the entire workload. For that, we configure MLPerf under its Offline inference workload for each DL application. To assure that the evaluation results are stable (\ie the confidence interval is negligible) and are not affected by the system factors, we run each workload 1,000 times and report the average time to complete it.

\subsection{Testbed Configuration}
As mentioned earlier, we evaluate four predominantly used execution platforms, namely Bare-Metal (BM), Virtual Machine (VM), Micro Virtual Machine (MVM), container in Bare-Metal (CN), and container in VM (VMCN). 
Table~\ref{table:config} elaborates on the specifications of each platform. The third column of the table shows the software and libraries within each execution platform that are installed inside two Anaconda environments (one with TensorFlow and another with PyTorch) to run the DL applications.  

\begin{table}[]
\hspace{-0.1in}
\begin{tabular}{c|c|c}
\hline
\textbf{Exec. Platform} &
  \textbf{Platform Specifications} &
  \textbf{DL Specs.} \\ \hline
BM &
  Ubuntu 18.04.3, Kernel 5.4.5 &
  \begin{tabular}[c]{@{}c@{}}Anaconda v 2020.11,\\ MLPerf v1.1,\end{tabular} \\ \cline{1-2}
VM &
  \begin{tabular}[c]{@{}c@{}}Qemu 2.11.1, Libvirt 4\\ Ubuntu 18.04.3, Kernel 5.4.5\end{tabular} &
  \begin{tabular}[c]{@{}c@{}}Python 3.8, \\ TensorFlow 2.4,\end{tabular} \\ \cline{1-2}
MVM &
  \begin{tabular}[c]{@{}c@{}}FireCracker 0.24.3, Libvirt 4\\ Ubuntu 18.04.3, Kernel 5.4.5\end{tabular} &
  \begin{tabular}[c]{@{}c@{}}PyTorch 1.5, \\ Cuda 10.2,\end{tabular} \\ \cline{1-2}
CN/VMCN &
  \begin{tabular}[c]{@{}c@{}}Docker 19.03.6, Ubuntu 18.04.3,\\ nvidia-docker2\end{tabular} &
  CuDNN 7.5 \\ \hline
\end{tabular}
\vspace{0.1in}
\caption{\small{Specifications of different execution platforms used in the evaluations.
The first column shows the acronyms we use for each execution platform. Libraries and packages that are used to evaluate the four DL applications offered by the MLPerf, which is the same across all the platforms, are in the third column.}}
\vspace{-0.2in}
\label{table:config}
\end{table}

To assure that the results are reproducible, and the execution platform is the only source of the overhead, we conduct the evaluations on an on-premise physical server as opposed to the public Cloud providers (\eg AWS and Azure) where the performance is inherently prone to randomness, due to issues like noisy neighbors and multi-tenancy \cite{potdar2020performance}.
The physical host used as our Cloud server is a DELL PowerEdge R840 with 2$\times$Intel Xeon Gold 6138 CPU with 80 homogeneous cores, 512 GB memory (DDR4 RDIMM 2666 MHz), and RAID1 (2$\times$900 GB HDD) storage. Each processor is 2.0 GHz with a 28 MB cache and 20 processing cores (40 threads). The GPU hardware is a NVIDIA Tesla Volta k40 with 11 GB of memory. The GPU is attached to the VMs using a \textit{Pass-through} approach, and to the containers using the NVIDIA Container Toolkit \cite{yang2012using}. A Raspberry Pi 4 device is used as the Edge testbed. It includes a Quad core Cortex-A72 (ARM v8) 64-bit SoC@ 1.5GHz, and 8 GB LPDDR4-3200 SDRAM with a 32GB Micro SDHC memory (80 MB/s transfer speed).

\subsection{Evaluation Scenarios}
\label{evaluationScenarios}
\begin{figure}[ht]
  \centering
  \includegraphics[width=0.50\textwidth]{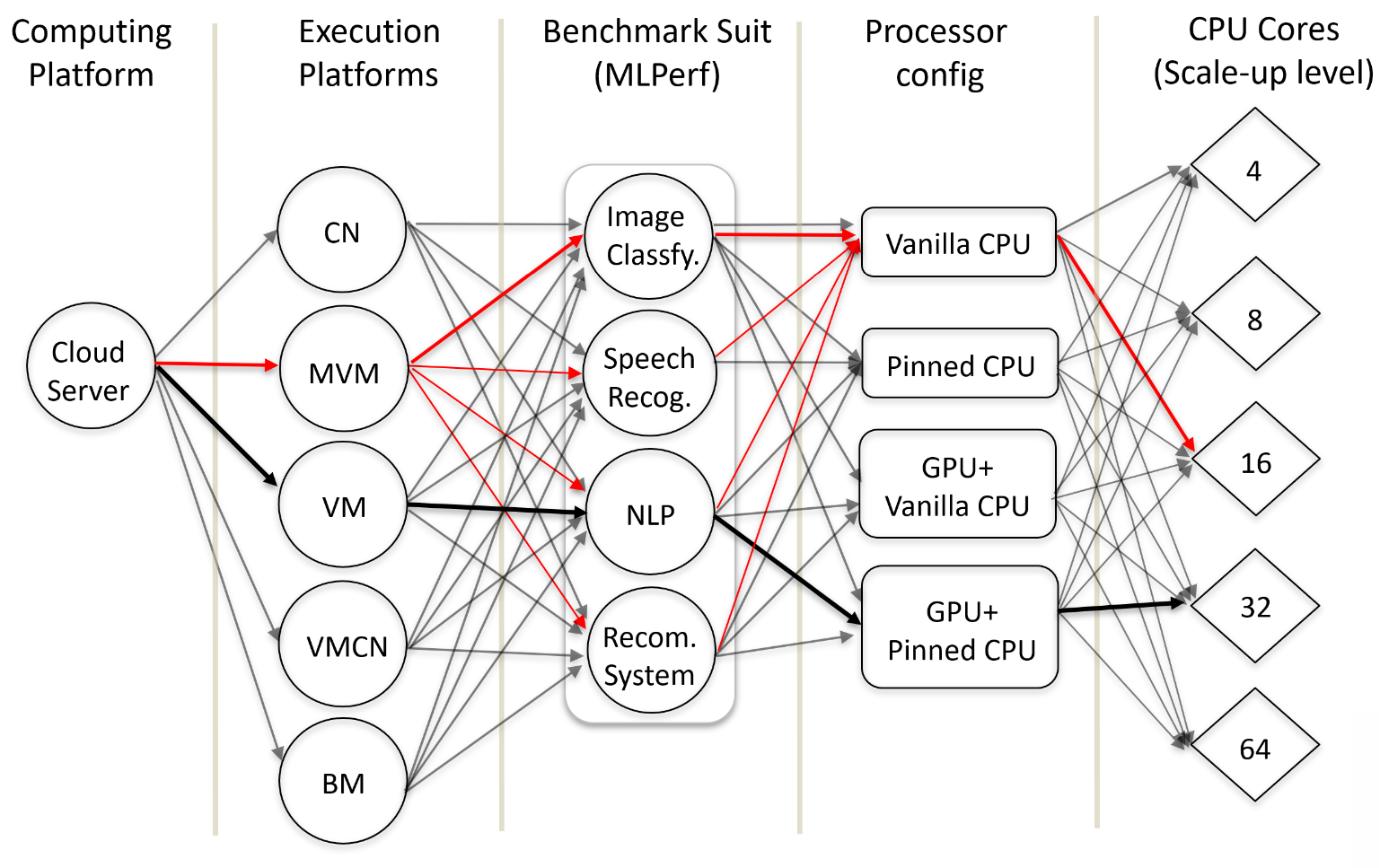}
  \caption{\small{A graph representing all the scenarios evaluated in this study. Each directed path from the beginning to the end shows one scenario. For instance, the solid black path indicates the scenario of a VM with NLP application, configured with 32 pinned CPU cores and the GPU. Due to the technical limitations of MVM (Firecracker), its scenarios are restricted to the ones represented by the red color.}}
  \label{Fig:scenarios}
   \vspace{-0.1in}
\end{figure}

Figure \ref{Fig:scenarios} expresses all the scenarios that we evaluate on the Cloud server. As we can see, the evaluations encompass different execution platforms, each one tested against different DL applications of MLPerf, under different processor types and configurations, and different number of cores (scale-up level). For the sake of simplicity, we use the following naming convention to identify each configuration: \textbf{exec. platform\textcolor{red}{/}scale-up level\textcolor{red}{/}CPU config.\textcolor{red}{/}processor type}. Options for the CPU configurations are \{\textbf{VNL, PIN}\} that refer to vanilla and pinning configuration, respectively. Options for the \emph{processor type} are \{\textbf{CPU,} \textbf{GPU}\} that refer to homogeneous CPU-based, and heterogeneous CPU-GPU systems, respectively. For instance, the solid black arrows show a scenario where the NLP application is deployed with the \textbf{VM/32/PIN/GPU} configuration (\ie a VM with 32 pinned CPU cores and GPU).

We note that features like CPU pinning and GPU pass-through are not yet supported on MVMs and Firecracker. Thus, for MVM, we only evaluate MVM/16/VNL/CPU configuration for various DL applications. In Figure \ref{Fig:scenarios}, red lines depict all the scenarios evaluated by the MVM. We dedicate one subsection (\cref{subsec:cmp}) to the analysis of this particular case. 

We also study the overhead imposed by different execution platforms when DL application are deployed on the Edge device. For the sake of clarity, these scenarios are not shown in Figure~\ref{Fig:scenarios}. For consistency across the evaluations, we configure the IC and SR applications with ResNet50 and RNNT as their DNN models. We apply the same inference workload that is used on the Cloud server. Due to the lack of elasticity (scale-up) and pinning support on the Edge, we only use the 4 cores of the Raspberry Pi without pinning.

\section{Application-Specific Analysis of the overhead}
\label{sec:applicationSpecific}
\subsection{Image Classification (IC)}\label{subsec:app}
In this part, we study the overhead of the execution platforms to perform E2E inference of the IC application. We first elaborate on the E2E inference in the Cloud-based server and then, we extend the analysis to DNN model inference time. 

\subsubsection{Analyzing the Overhead on Homogeneous and Heterogeneous Cloud Servers}
\begin{figure*}[] 
\vspace{-0.1in}
\centering
\hspace{-0.3in}
\subfloat[Homogeneous CPU-based system]{
\includegraphics[width=0.45\textwidth]{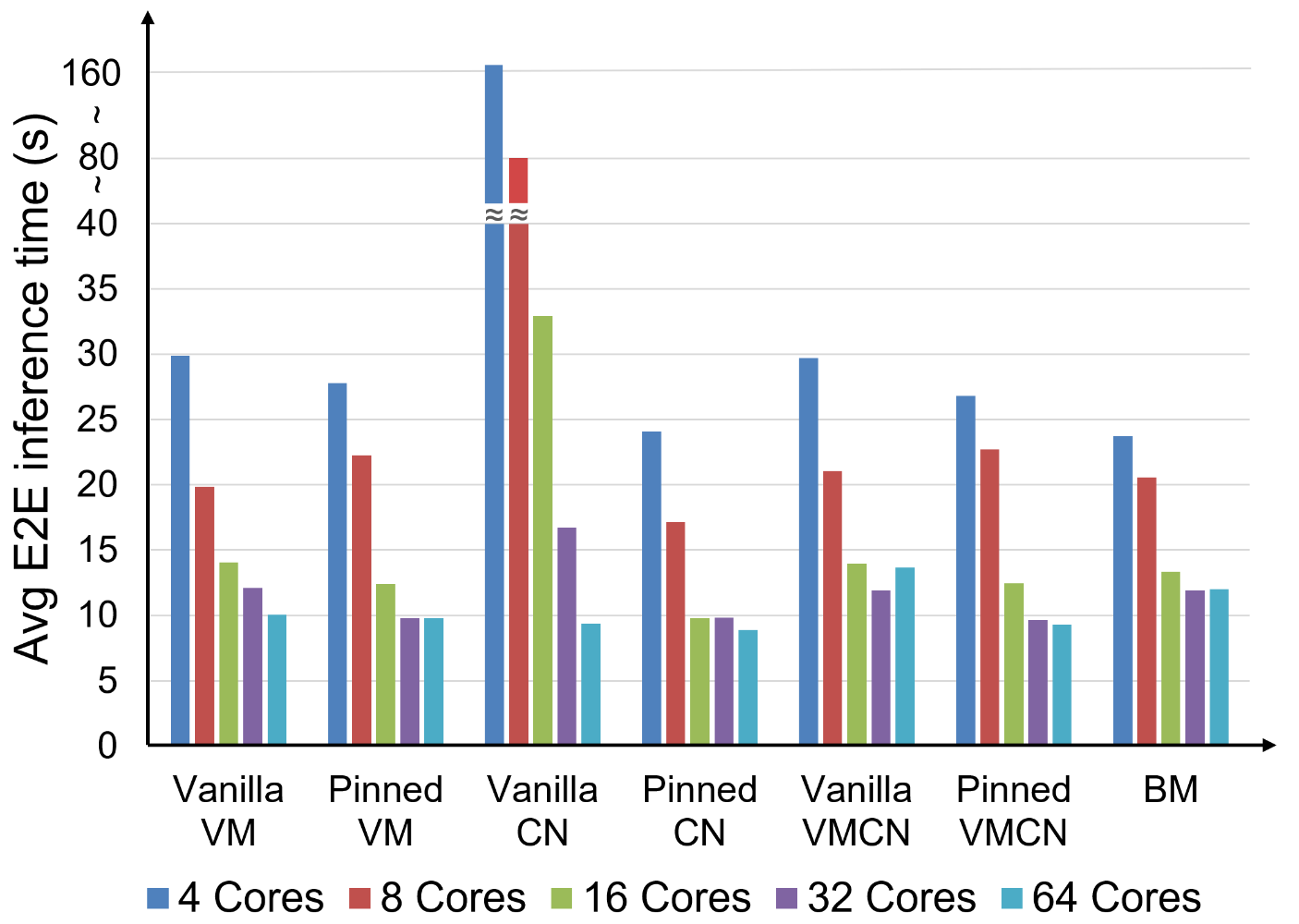} 
\label{fig:visionCPU}
}
\hspace{0.4in}
\subfloat[Heterogeneous CPU-GPU system]{
\includegraphics[width=0.45\textwidth]{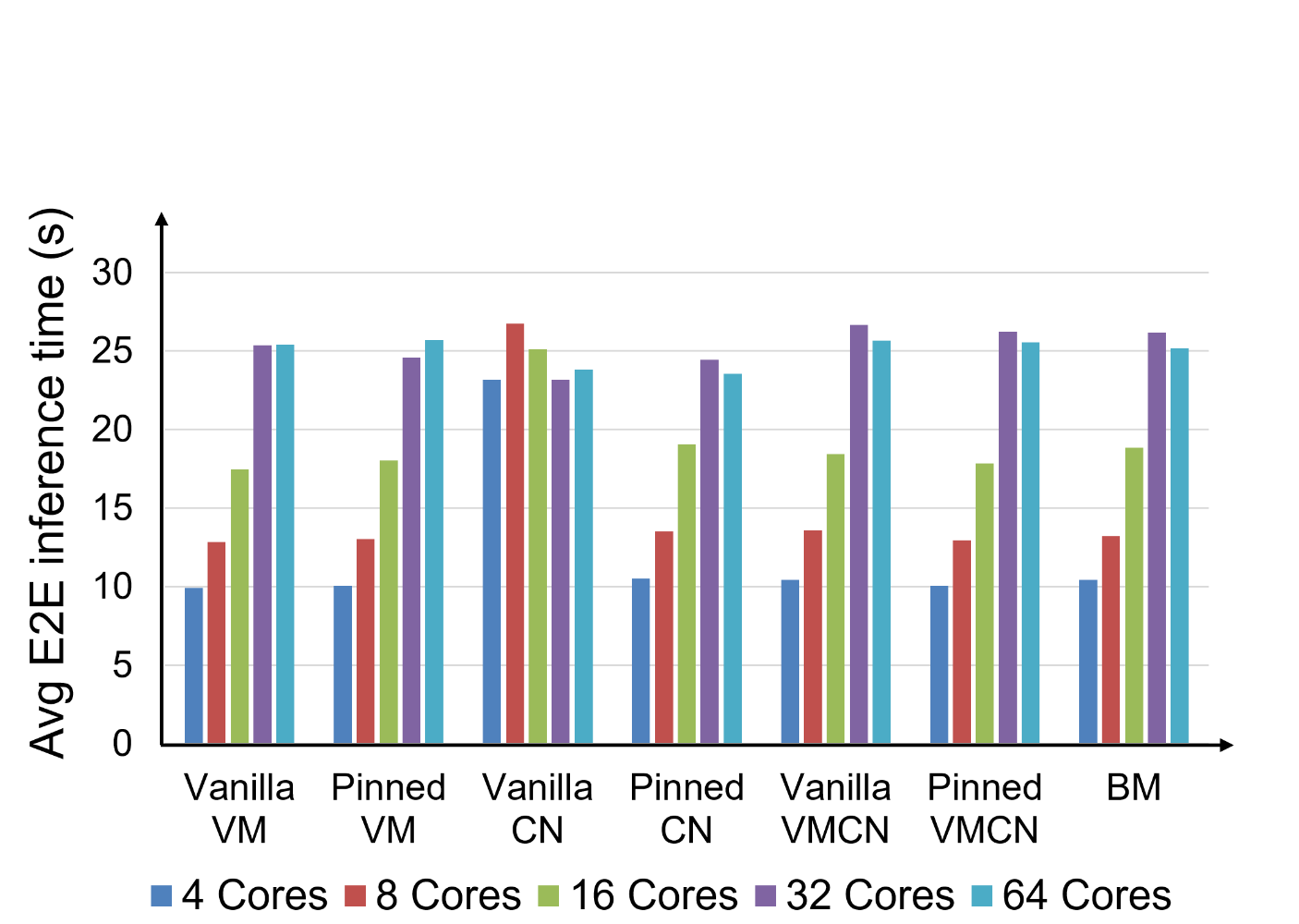} 
\label{fig:visionGPU}
}

\caption{\small{Average E2E inference time (in Seconds) of the image classification application in different execution platforms, when the scale-up level ranges from 4 to 64 CPU cores in each group of bars. The input workload workload characteristics are provided in Table \ref{table:AppCharacteristics}.}}
\vspace{-0.1in}
\label{fig:vision}
\end{figure*}

The evaluation results of the IC application on the homogeneous (CPU-based) Cloud server are depicted in Figure \ref{fig:visionCPU}. The horizontal axis in this figure shows the scale-up for various execution platforms, and the vertical axis shows the average E2E inference time of each configuration. In this figure, we can see that, in general, scale-up improves the E2E inference time up to a certain number of cores. However, due to the  Amdahl's law, there is no major improvement for scale-ups beyond 32 cores. We can see that in this application, regardless of the execution platform, scale-up level 16 can provide an acceptable cost/performance trade-off. 

While containers (particularly, Vanilla CN and Vanilla VMCN) are expected to impose the least overhead across all the virtualization platforms, in deployments with less than 16 cores, we observe the opposite behavior. However, this spike overhead fades away upon scale-up beyond 16 cores. As this behavior is observed repeatedly across all the applications, we defer the root cause investigation of it to the cross-application analysis section (see \cref{subsec:unusualCNbehavior}).  

Comparing the E2E inference times across Figures \ref{fig:visionCPU} and \ref{fig:visionGPU} demonstrates that utilizing CPU-GPU platforms is not remarkably beneficial for this application. Monitoring utilization of the execution platforms revealed that this is because it is the pre- and post-processing phases of the application on the CPU that form the bottleneck, and only the inference phase uses GPU and it constitutes just a fraction of the E2E inference time. That is why, in this particular application, utilizing GPU does not effectively enhance the E2E inference time. 
Figure \ref{fig:visionGPU} also shows that, regardless of the execution platform, unexpectedly, the E2E inference time substantially rises upon increasing the scale-up level. The reasons for this counterintuitive behavior in the IC application are as follows:
\begin{enumerate}[label=(\alph*)]
     \item The complexity and overhead of data transfer between the main memory and the GPU memory increases upon scale-up. In fact, with more cores and threads, the contention for copying data to the GPU memory is amplified. 
     \item The number of cache misses is increased. One reason for this is splitting the E2E inference across the CPU (for pre- and post-processing) and GPU (for the inference) that increases the number of CPU interrupts, and subsequently, the number of cache misses. Another reason is that, in the IC application, the pre- and post-processing phases are not developed to effectively make use of excessive number of cores (\ie their degree of parallelism is limited). In this situation, scale-up can even aggravate the performance, because it leads to context switch to other CPU cores that causes more cache misses \cite{liu2010understanding}.

\end{enumerate}

\subsubsection{Verifying the Impact of Pre- and Post-Processing Phases on the Speedup}
To assure that, when GPU is used, the pre- and post-processing phases are the reasons for the slow down in the E2E inference, we conduct another experiment in which the speedup resulted from the E2E inference is compared against the speedup of the inference-only operation (\ie when only the DNN inference is performed and the pre- and post-processing phases are excluded). For that purpose, we employ DLBS to benchmark the ResNet50 model inference. We configure both the CPU-based and GPU-based execution platforms with 16 cores, and feed them with the same IC workload. For a given execution platform, we define the \emph{speedup} as the ratio of execution time on the CPU-based platform to the execution time on the GPU-based one.

\begin{figure}[ht]
  \centering
  \vspace{-0.1in}
  \includegraphics[width=0.5\textwidth]{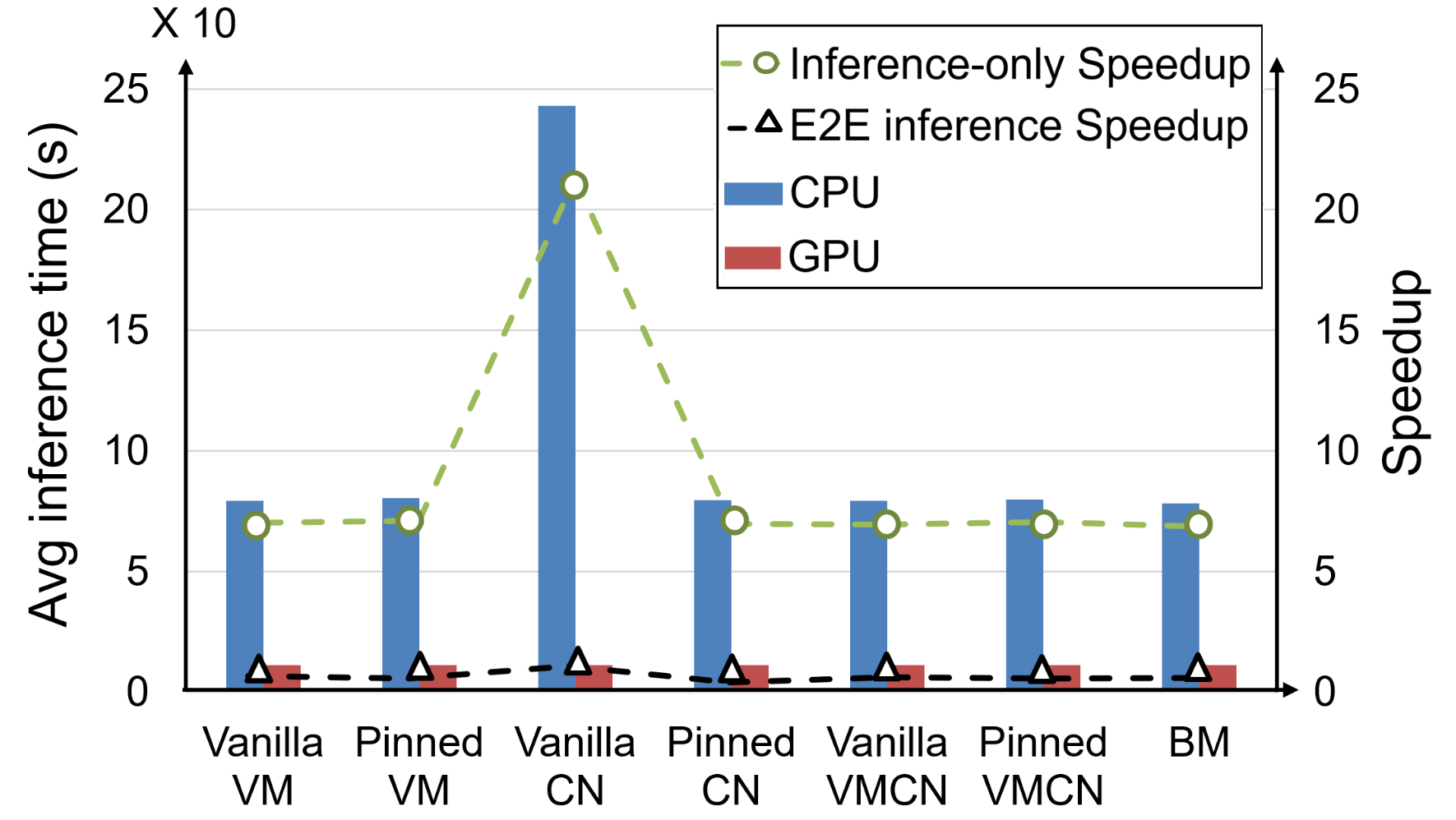}
  \caption{\small{Benchmarking the ResNet50 DNN model inference on the CPU-based and GPU-based execution platforms. The left-side vertical axis shows the average inference time and the right-side vertical axis shows the speedup of using GPU versus CPU both for the inference-only and E2E inference operations.}}
  \label{Fig:DLBS}
  \vspace{-0.2in}
\end{figure}

The results of this evaluation, in Figure~\ref{Fig:DLBS}, demonstrate that configuring GPU-based platforms for the inference-only operation is effective and yields 4$\times$---9$\times$ speedup, as opposed to the situation where CPU-based platforms are used for the inference. Moreover, this figure demonstrates the E2E inference time speedup in the 16-core execution platforms. The remarkable drop we observe in the speedup is the impact of including pre- and post-processing phases. It is noteworthy that we evaluated other DNN models (particularly, ResNet18, 50, 152, and AlxNet) on the execution platforms and verified that our inference-only analysis is not DNN model-specific.

Provided that, in practice, pre- and post-processing are indispensable parts of DL applications, and considering the high price of GPU-based platforms in the public Clouds, we can state that the gain of using GPU for the IC application is only when the CPU scale-up level is low (less than eight). Importantly, we can conclude that higher scale-up levels are even counter-productive and can slow down the E2E inference.

\subsection{Speech Recognition (SR)}
\begin{figure}[ht]
\vspace{-0.1in}
  \centering
  \hspace{-0.1in}
  \includegraphics[width=0.50\textwidth]{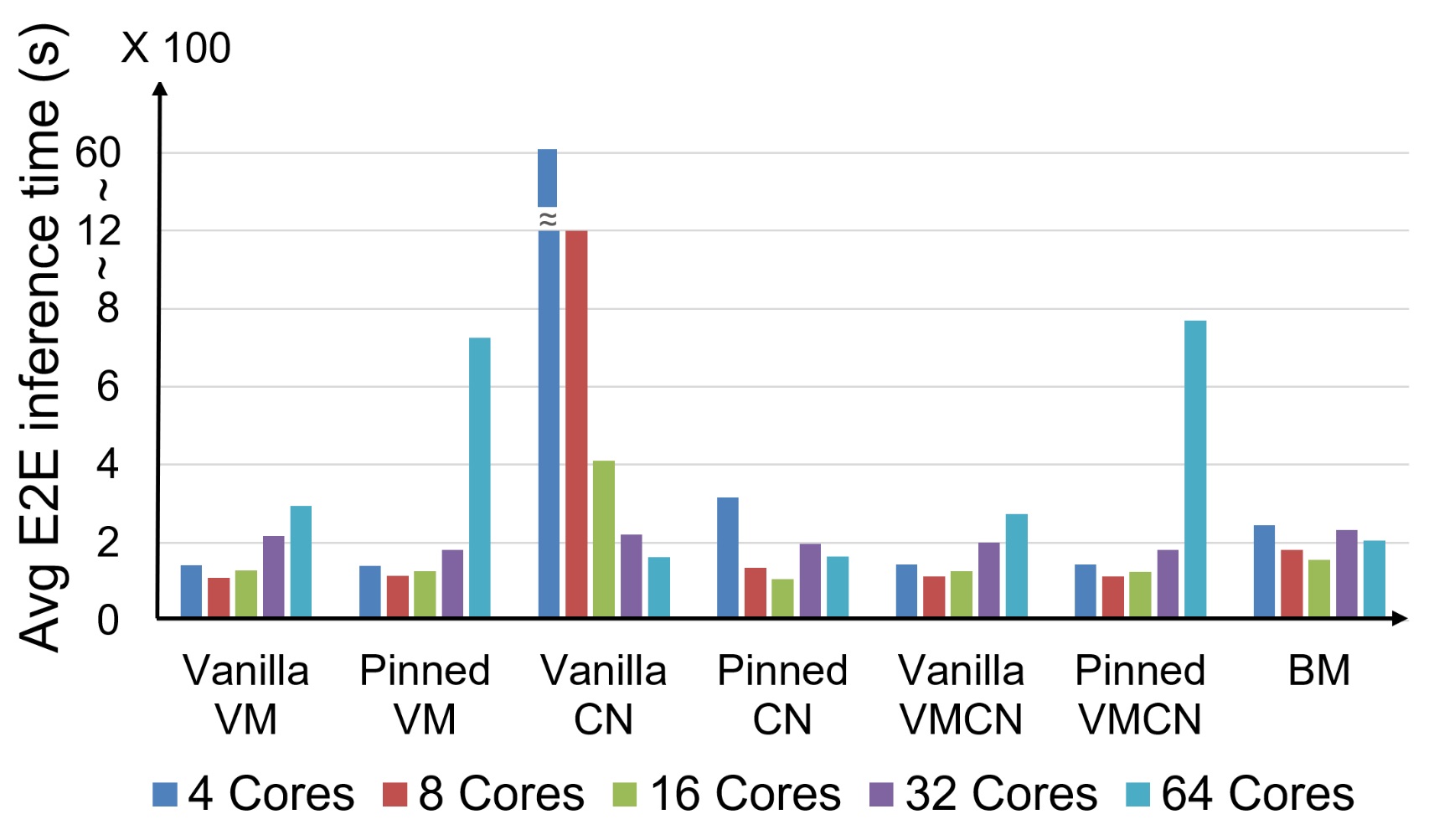}
  \caption{\small{Average E2E inference time of running the SR application on the CPU-based Cloud server. The horizontal axis shows the various execution platforms that are scaled-up from 4 to 64 CPU cores.}}
  \label{Fig:SpeechCPU}
   \vspace{-0.2in}
\end{figure}

Recall from Table~\ref{table:AppCharacteristics} that the SR application in MLPerf is incompatible with GPU, thus, in Figure \ref{Fig:SpeechCPU}, we report the E2E inference time only for the homogeneous CPU-based configurations. In this figure, we observe that, except vanilla CN (that is discussed later in section \ref{subsec:unusualCNbehavior}), other virtualization platforms have a similar trend for the scale-up level up to 32 cores. However, in the VM-based platforms with 64 cores (particularly, the pinned ones: VM/64/PIN/CPU and VMCN/64/PIN/CPU), the E2E inference time has skyrocketed. This is because on our server, 64-core instances are the only ones allocated across two CPU sockets. Communication across these sockets via VMs is known to impose a significant overhead \cite{majo2011memory}. That is why a similar behavior is observed on 64-core VM-based instances in Figures \ref{fig:NLPCPU} and \ref{fig:RecommendationPU}.

A closer observation shows that the E2E inference time is reduced upon scale-up to 8 CPU cores, however, it is deflected for the higher scale-up levels. Since this trend is repeated regardless of the underlying execution platform (except vanilla CN), we attribute it to the way the application's multi-threading is developed. Because a similar behavior is observed in other DL applications, we defer further analysis of this to the cross-application analysis section (\cref{subsec:paralelization}). 

\subsection{Natural Language Process (NLP)}

\begin{figure*}[] 
\vspace{-0.1in}
\centering
\subfloat[Homogeneous CPU-based system]{
\includegraphics[width=0.45\textwidth]{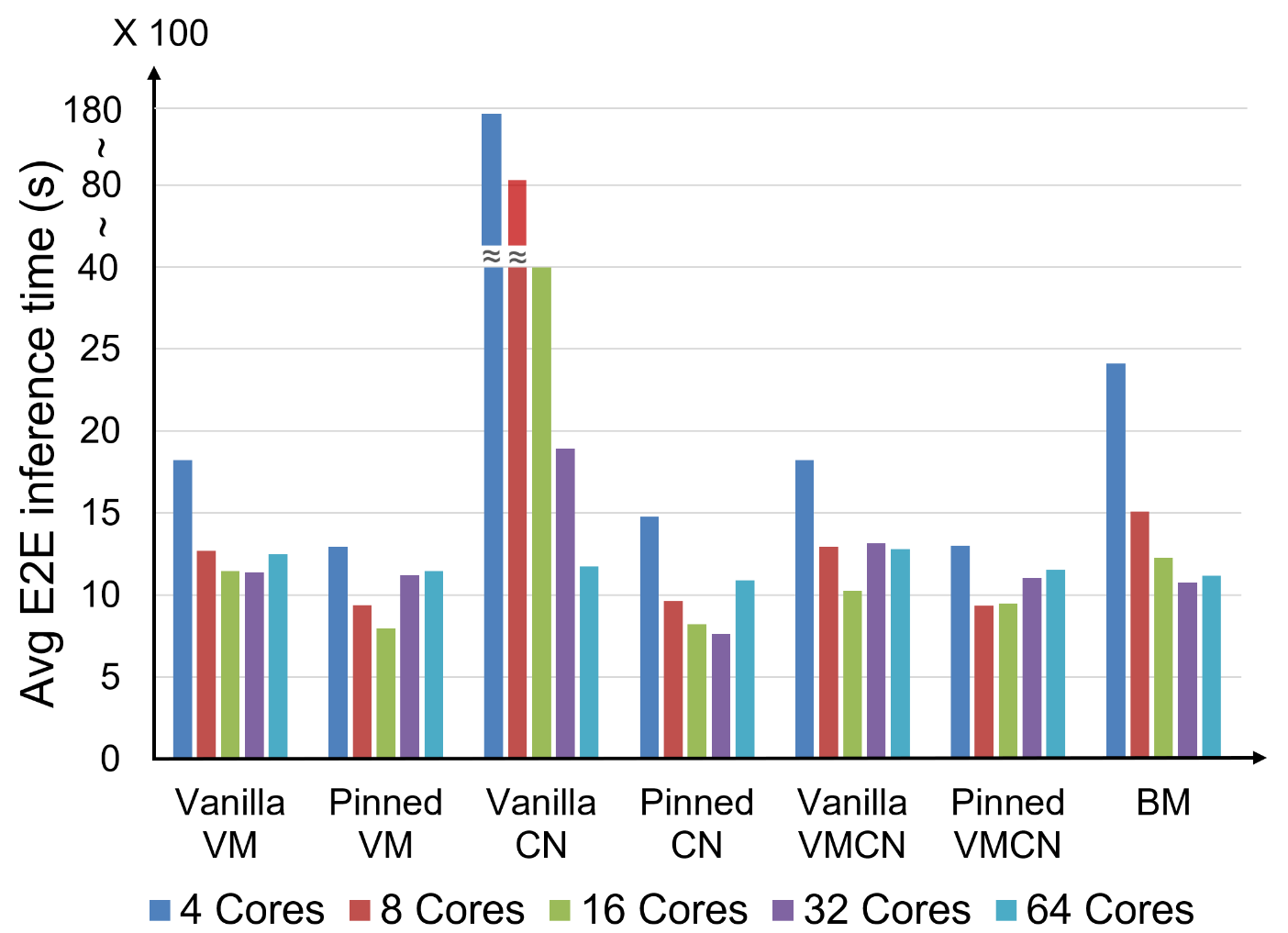} 
\label{fig:NLPCPU}
}
\subfloat[Heterogeneous CPU-GPU system]{
\includegraphics[width=0.45\textwidth]{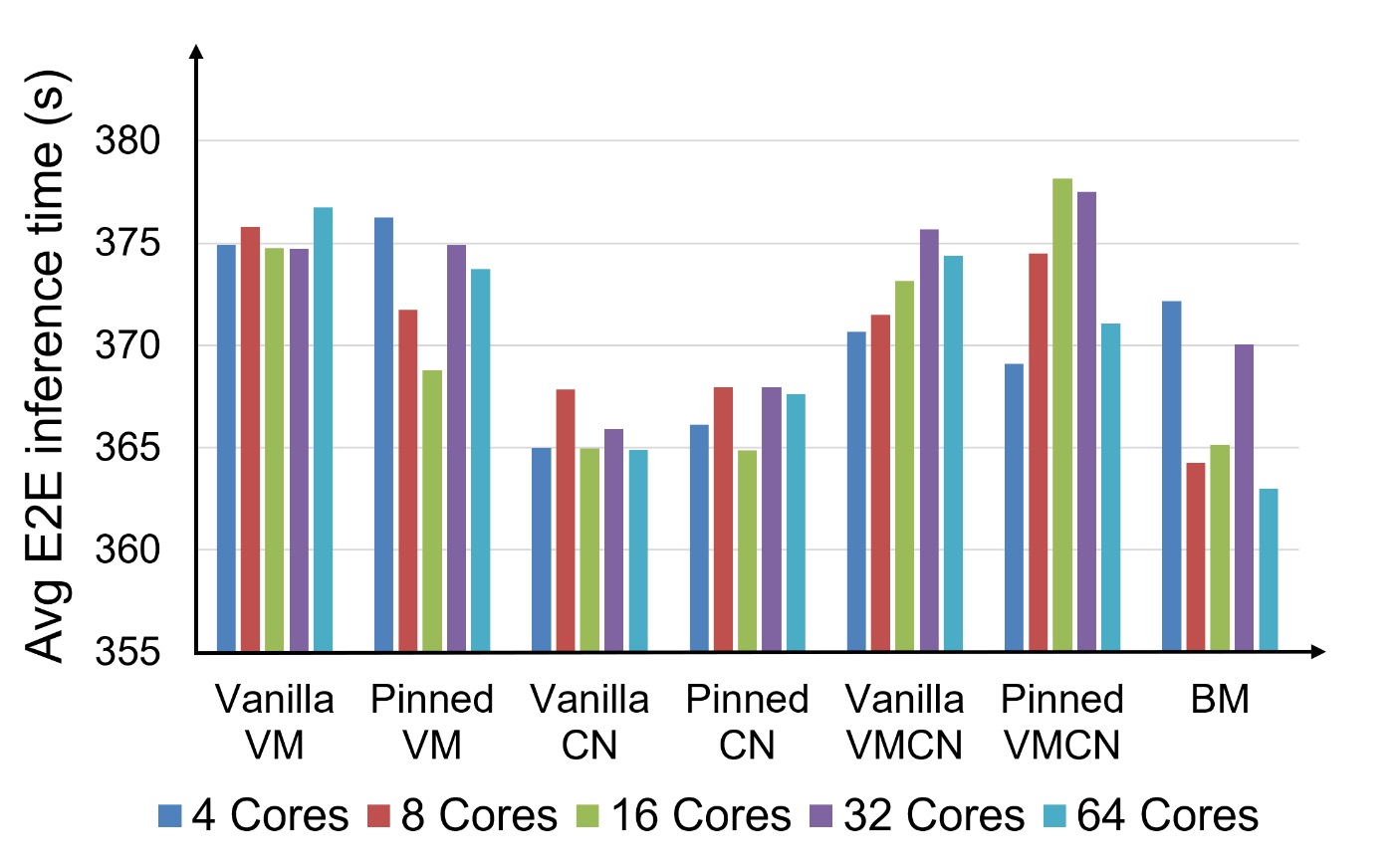} 
\label{fig:NLPGPU}
}
\vspace{-0.07in}
\caption{\small{Average E2E inference time of the NLP application across various execution platforms. The scale-up level ranges from 4--64 cores.}}
\label{fig:NLP}
\end{figure*}

Analysis of the NLP application is conducted on both homogeneous (CPU) and heterogeneous (CPU-GPU) systems. For the CPU case, shown in Figure~\ref{fig:NLPCPU}, scale-up level 16 offers the lowest E2E inference time across all the virtualization platforms (excluding vanilla CN). Particularly, pinned platforms (pinned VM, CN, and VMCN) effectively reduce the overhead and offer the minimum E2E inference time in compare to vanilla ones. The justification for the positive impact of pinning and negative impact of scale-up beyond 16 cores are described in the cross-application section (\cref{sec:crossAnalysis}).

The NLP application on CPU-GPU configuration expresses a different behavior. As shown in Figure \ref{fig:NLPGPU}, there is a remarkable improvement in the E2E inference time as opposed to the CPU configuration in Figure \ref{fig:NLPCPU}. In all of the execution platforms, GPU causes more than 3$\times$ speedup in the E2E execution time. This is because, unlike the IC application, in NLP, the inference time on GPU dominates the times of pre- and post-processing on CPU. For the same reason, the scale-up level and pinning are not a decisive factors in this case and a 4-core unpinned platform offers a satisfactory performance.

\vspace{-0.04in}
\subsection{Recommendation Systems (RS)}
\begin{figure*}[] 
\centering
\vspace{-0.1in}
\subfloat[Homogeneous CPU-based system]{
\includegraphics[width=0.43\textwidth]{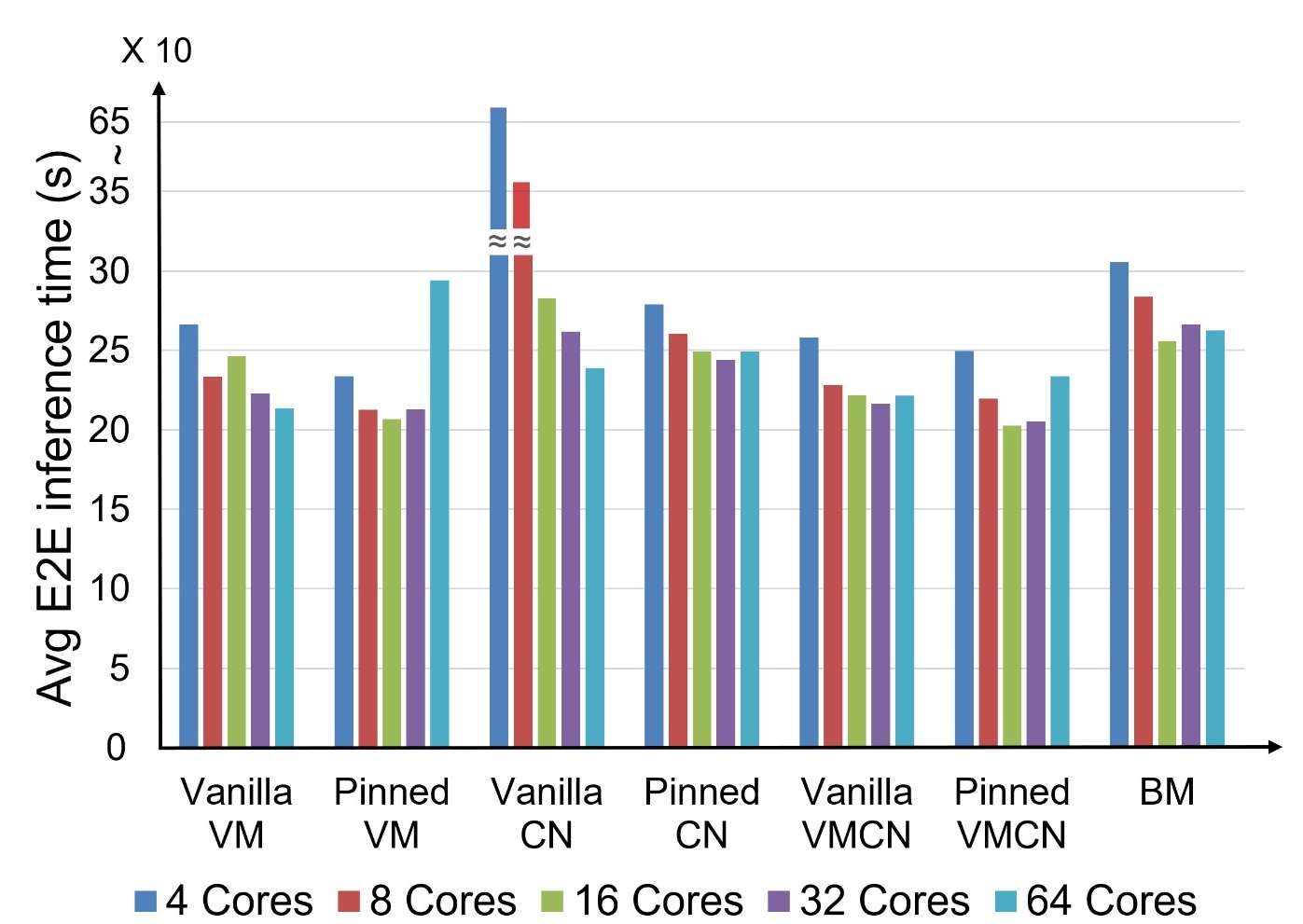} 
\label{fig:RecommendationPU}
}
\hspace{0.5in}
\subfloat[Heterogeneous CPU-GPU system]{
\includegraphics[width=0.35\textwidth]{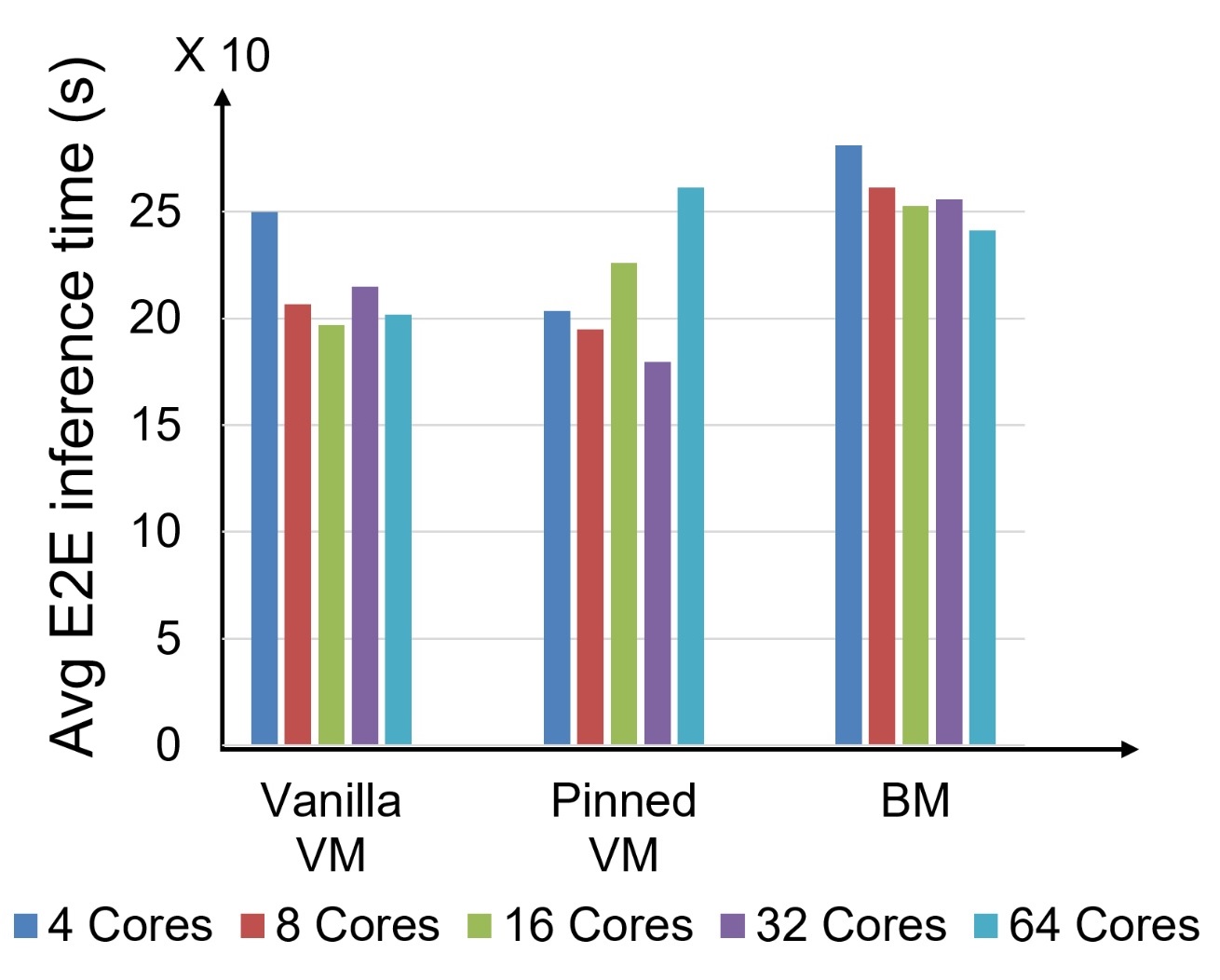} 
\label{fig:RecommendationGPU}
}

\vspace{-0.07in}
\caption{\small{Average E2E inference time of the Recommendation Systems (RS) application in Seconds (s). The horizontal axes are grouped by the execution platforms, and within each group, the platform is scaled-up from 4 to 64 CPU cores. Because RS is incompatible with GPU under the containerized platforms, the container-based platforms are excluded in the subfigure (b). }}
\vspace{-0.1in}
\label{fig:Recommendation}
\end{figure*}

This application is run with both CPU and GPU on the Cloud server, however, due to the incompatibility with the GPU hardware on the Docker container, the containerized platforms are excluded from the evaluation in Figure~\ref{fig:RecommendationGPU}.

Analyzing the results on the homogeneous CPU-based system, shown in Figure \ref{fig:RecommendationPU}, expresses the following facts that confirm our observations for other applications: (a) the outlier overhead of vanilla CN; (b) no improvement by scale-up to 64 cores; and (c) the positive impact of CPU pinning. These are all unraveled in the cross-application analysis Section~\ref{sec:crossAnalysis}. 

Comparing the results of the CPU-based system against the GPU-based one (Figure~\ref{fig:RecommendationGPU}) reveals that GPU does not significantly improve the E2E inference time of the RS application. Similar to the IC application, this is because the inference operation (that utilizes GPU) constitutes only a fraction of the E2E inference time, whereas, the pre- and post-processing phases (that utilize CPU) dominate the E2E inference time. This evidences suggest Cloud solution architects to study/profile the DL application behavior, before lavishly utilize expensive---and sometimes inefficient---GPU instances. 

\section{Cross-Application Analysis of the Overhead}
\label{sec:crossAnalysis}
\subsection{Why the Overhead of Vanilla CN is an Outlier?} \label{subsec:unusualCNbehavior}
To maximize revenue, Cloud providers deploy large servers (with many CPU cores) in their data centers that encourages multi-tenancy and better use of the shared resources (\eg memory and storage). In this large servers, the size of user virtual instances are, often, substantially smaller than the number of server (a.k.a. physical host) CPU cores. Container to Host Ratio (a.k.a. \emph{CHR}) is defined as the ratio of number of container CPU cores to the number of CPU cores in the server. It is proven that the value of CHR has an inverse impact on the overhead of the vanilla CN  \cite{samani2020art}. That is, the lower values of CHR lead to a higher overhead and vice versa. The reasons for such a counter-intuitive behavior in vanilla CN are as follows: 

\paragraph{The host OS scheduler} For vanilla CN, the host OS scheduler is the ultimate decision maker that allocates CPU cores to the container. The scheduler potentially assigns a different set of cores to a vanilla CN at each time slot. This implies migrating processes from one set of cores to another at each time slot that is expected to induce an overhead due to the frequent context switch, cache miss, and reestablishing interrupts for IO operations \cite{wong2008fairness}. For a small vanilla CN (\ie with a low CHR value), these implications are more influential, because there are more possible destination cores.  
\paragraph{Control Groups} (a.k.a. \texttt{cgroups}) that is in charge of accounting and policing resource usage of containers \cite{Josef}. This kernel module has to assure that the cumulative CPU usage of the vanilla CN does not exceed its designated quota. In each scheduling event, a vanilla CN undergoes the overhead of both OS scheduling (that implies process migration) and cgroups (for resource usage tracking). In particular, because cgroups is an atomic process, the container has to be suspended to switch between user-mode and kernel-mode, until tracking and aggregating its resource usage is complete. 

\subsection{Why Bare-Metal Overhead is Higher than the Virtualization Platforms?}
\begin{figure}[ht]
\vspace{-0.18in}
  \centering
  \includegraphics[width=0.4\textwidth]{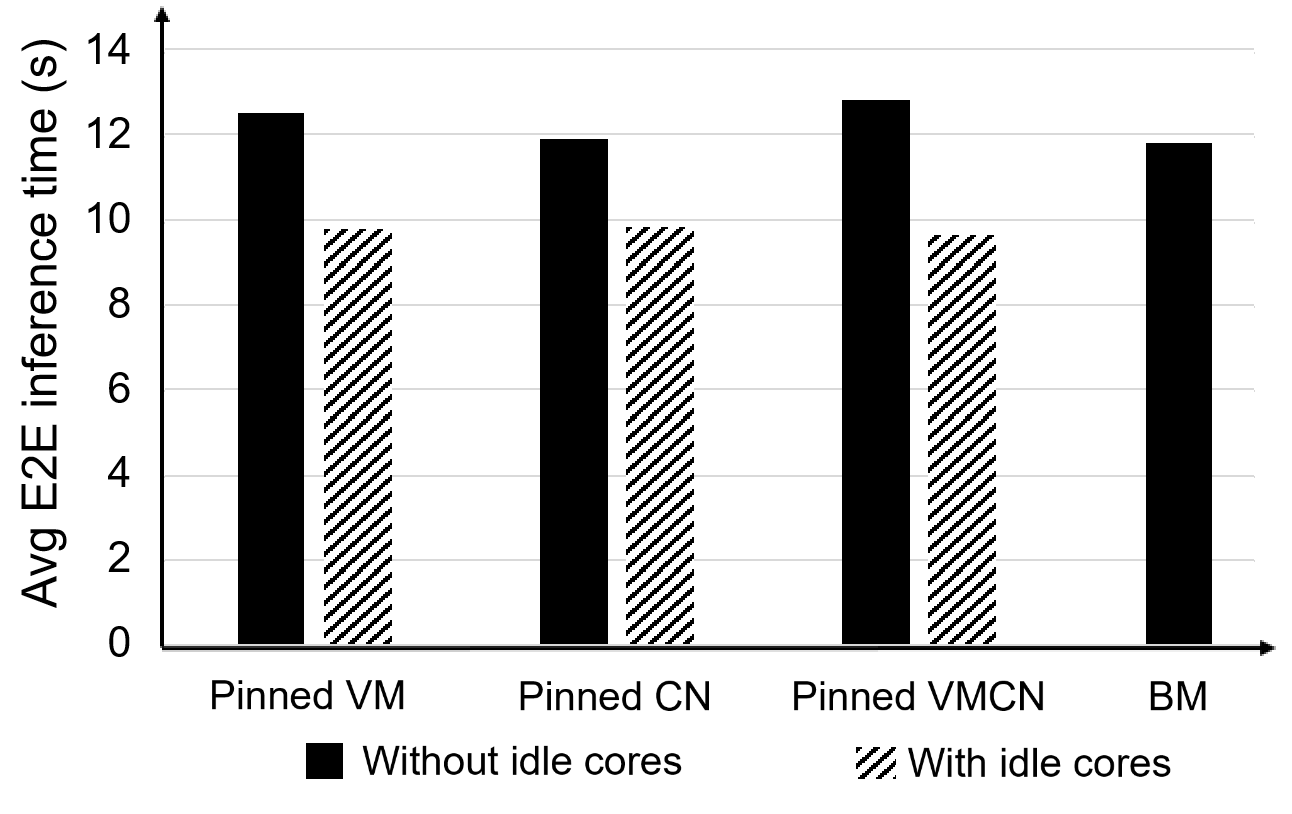}
  \vspace{-0.1in}
  \caption{\small{Average E2E inference time of the IC application for different platforms with 32 cores on a server with 32 CPU cores (no idle core).}}
  \label{Fig:OddBM}
   \vspace{-0.14in}
\end{figure}

Despite our expectations, we observed that BM fails to impose the lowest overhead in several cases. For instance, in Figure \ref{fig:visionCPU}, we see that for the scale-up level of 32 and higher, the E2E inference time is higher than the virtualization platforms. The reason behind this behavior is related to the way virtualized platforms are provisioned. Practically, in datacenters, VMs and Containers are instantiated in a host with numerous CPU cores, whereas, in BM there is no such idle core (\ie all allocated cores are utilized for processing). For example, in this study, the VM with 32 CPU cores is created on a host with 80 CPU cores. While the guest OS has access to its own 32 cores, the host OS can utilize the rest of the CPU cores for other system processes. This implies that, having idle CPU cores on the host mitigates the resource contention between the guest and host OSs, thereby, reducing the overhead of the virtualized platform. To verify this hypothesis, we conducted an experiment with 32-core CN, VM, and VMCN platforms, on a host with 32 CPU cores and configured the IC application as the workload. The results, depicted in Figure \ref{Fig:OddBM}, show that without idle CPU cores on the host OS, the overhead of the virtualization platforms (particularly, vanilla VM and VMCN) are significantly higher than BM. 

\subsection{The Impact of CPU Pinning on the Platform Overhead}
In contrast to the vanilla mode, CPU pinning avoids the overhead of the host OS's scheduling and \texttt{cgroups} discussed earlier. In the presence of pinning, as the allocated set of processing cores does not change at each scheduling event, there is less demand for \texttt{cgroups} invocation, hence, the imposed overhead is mitigated. Based on the results, pinning is advantageous in other virtualization platforms, and solution architects should use them for DL application deployments.

\subsection{The Impact of Parallelization and Development Techniques on the Overhead}\label{subsec:paralelization}
Multi-threading is the common technique developers employ to parallelize DL applications. However, the way multi-threading is implemented is highly dependent on the developer's expertise, and it is one of the important factors to assess before deploying the application in the production environment. 
On the one hand, in a CPU-GPU deployment (\eg the IC application in Figure \ref{fig:visionGPU}), because pre- and post-processing phases become the bottleneck, it is important that the developers efficiently parallelize them to break the bottleneck. On the other hand, to reduce the E2E inference time in a homogeneous CPU-based deployment, all the three phases of the E2E inference, and the external libraries used by these phases, should be multi-threaded. For example, in the case of IC application (Figure \ref{fig:visionCPU}), multi-threading is used in: \texttt{Numpy} and \texttt{OpenCV} pre-processing libraries, \texttt{tf.config.threading} as the TensorFlow DNN inference library, and \texttt{matplotlib} as the post-processing library. 

\subsection{The Impact of Input Type on the Platform Overhead}
Data types such as image or recorded audio files are large in size, whereas, other data types such as text files are often of small. From the storage  device perspective, the former data types demand a throughput-based storage device like HDD, whereas, the latter one desires a device with a high number of I/O operation per second (IOPS-based) like Solid-State Drives (SSD) that can serve up to 100$\times$ more IOPS than HDDs. 

The impact of storage subsystem on the E2E inference time is particularly substantial in the CPU-GPU system, because in this case, pre- and post-processing phases that directly perform I/O operations dominate the E2E inference time, hence, improving them is directly translated to the improvement in the E2E inference time. 
Provided that we use HDD in our evaluations, we can see that for the throughput-based input types (\eg the IC application in Figure \ref{fig:visionGPU}), there is not any major performance difference between CN and VM/VMCN platforms. In contrast, for the IOPs-based input types (\eg NLP in Figure \ref{fig:NLPGPU}), we observe a performance different across the same virtualization platforms. This discrepancy in the overhead across the CN-based versus VM-based platforms is because of using the inefficient (non-IOPS-based) storage device and is aggravated by multiple abstraction layers in the VM-based platforms. Upon increasing the scale-up level and, subsequently, the number of concurrent threads, there is more IOPS demand that further slows down the E2E inference.

\subsection{The Impact of Scale-up Level WRT Size of Physical Host}
One observation across the evaluation results of all DL applications is that, regardless of virtualization platform, scale-up does not necessarily lead to a lower E2E inference time. This pattern is specifically evident for the higher scale-up levels, \eg 64 CPU cores in Figures \ref{fig:visionGPU} and \ref{Fig:SpeechCPU}. While the slow down is partly due to the Amdahl's law, the fact that for the instances with numerous CPU cores, more than one CPU socket is utilized also contributes to this behavior. In particular, interconnection across CPU sockets and inability to use NUMA deteriorate the processing and memory access latencies, thereby, increasing the E2E inference time. Our Cloud server utilizes two 40-core CPU sockets. Scale-up levels less than 32 enjoy on-chip communications and NUMA (\ie fast local memory access), whereas, those beyond 40 do not. 

Another across-application analysis with regards to the scale-up show that, for the evaluated applications, irrespective of their underlying execution platform and configuration, scale-up level 16 offers the best cost/performance trade-off. 

\subsection{Analysis of the Micro VM (MVM) across DL Applications}\label{subsec:cmp}
\begin{figure}[ht]
  \centering
   \vspace{-0.14in}
  \includegraphics[width=0.5\textwidth]{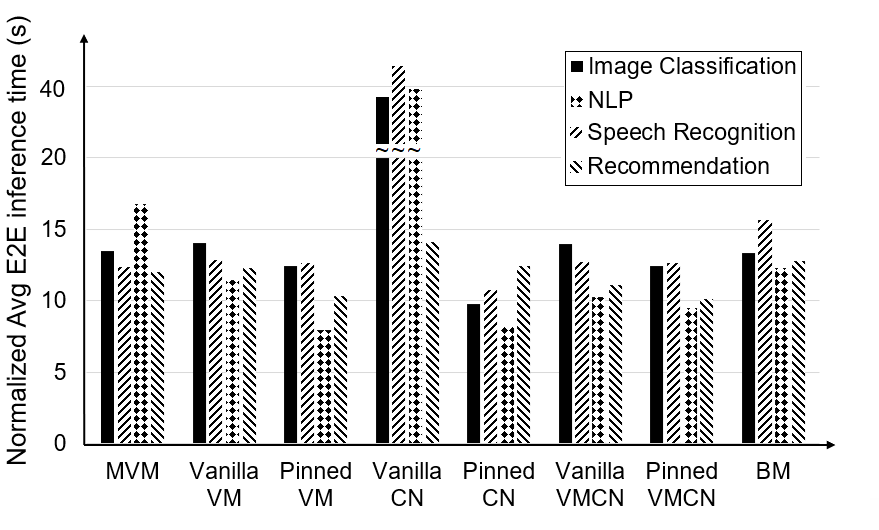}
   \vspace{-0.23in}
  \caption{\small{Average E2E inference time of DL applications for the MVM/16/VNL/CPU and other platforms with the same configuration.}}
  \label{Fig:DLApps-FC}
\end{figure}

For the MVM platform, we explored the performance of the MVM/16/VNL/CPU and MVM/32/VNL/CPU configurations for different DL applications and compared them against other execution platforms. However, due to the similarity in the results and shortage of space, in Figure~\ref{Fig:DLApps-FC}, we only report the results of 16 cores. In the figure, we normalized the average E2E inference times of applications to bring them to the same scale and make them presentable in one chart.

When comparing MVM with vanilla VM, we observed that, although MVM boots remarkably faster than vanilla VM, the difference in their E2E inference time is practically insignificant across all applications, except NLP. Even though MVM has less abstraction layers and is expected to improve the E2E inference time, because it lacks the Qemu para-virtualization modules that improve the IO performance (particularly, \texttt{virtio} that accelerates VM's disk access), the storage access latency is considerably increased. It is this increase in the IO latency that causes MVM either does not offer performance superiority over vanilla VM, or even increases the E2E inference time (as it is the case for NLP). To verify this hypothesis, we conducted another experiment using \texttt{Cassandra} \cite{shirinbab2017performance}, which is a highly IO-bound NoSQL application and can stress the IO performance of the MVM. The result (not shown due to the shortage of space) demonstrated that VM remarkably outperforms MVM, because of having specific modules to improve the IO performance.

\subsection{Analyzing the Overhead of Execution Platforms on Edge}
\label{subsec:edge}
To explore the overhead imposed by different execution platforms on the Edge, we deployed IC and SR applications on the Raspberry Pi 4 device. Since MLPerf does not provide RS for the Edge device, it is not considered in this experiment. Moreover, because the transformer library was incompatible with Ubuntu on the ARM architecture, NLP is also excluded from this experiment. Similar to the previous experiment, we normalized the average E2E inference times of the applications to make them chartable.

\begin{figure}[ht]
  \centering
  \vspace{-0.1in}
  \includegraphics[width=0.4\textwidth]{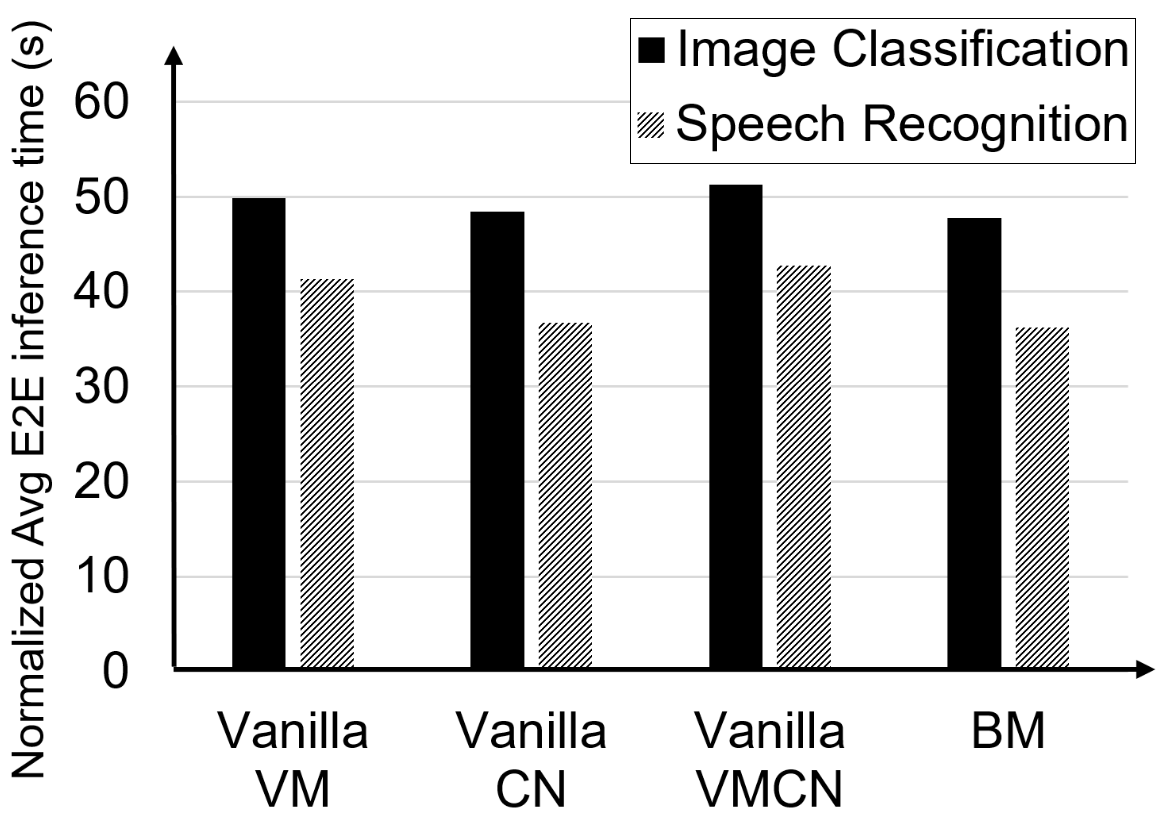}
  \caption{\small{Normalized average E2E inference time of the IC and SR applications (vertical axis) on different execution platforms deployed on the Edge. As CPU pinning is not possible on the Edge, only the vanilla cases are considered.}}
  \label{Fig:Edge Image Classification}
  \vspace{-0.1in}
\end{figure}

The result of this evaluation, in Figure \ref{Fig:Edge Image Classification}, shows that the CN platform introduces the lowest overhead (around 1 second more than BM) to process the IC application workload. The same behavior is exhibited for the SR application. The low overhead of CN is in contrast to the behavior of CN in the Cloud-based deployments. This discrepancy is because, unlike Cloud-based deployments, the number of cores assigned to CN on the Edge is the same as the number of cores that the device has (\ie CHR=1) that is proven to reduce the overhead \cite{samani2020art}. In addition, for both applications the imposed overhead rises for the VMCN platform, due to its multiple abstraction layers. In sum, CN is suggested as the suitable virtualization platform for DL deployments on the Edge. Making use of CN (instead of BM) can increase the isolation and mitigate the maintenance burden of Edge devices without significantly affecting the usability of the DL applications running on it.

\section{Summary and best practices}
\label{sec:summary}
In this study, we studied how the imposed overhead of different execution platforms offered by the Cloud providers can affect the usability of the DL applications in terms of their E2E inference time.
To that end, we measured the overhead of different execution platforms (BM, CN, VM, VMCN, and MVM) on the E2E inference time of DL applications using MLPerf benchmarking applications, namely Image Classification, Speech Recognition, NLP, and Recommendation Systems. In addition, we explored the impact of scale-up and heterogeneity (CPU-GPU) configurations under two CPU provisioning methods (vanilla and pinned CPU cores). The study revealed that: \textbf{(A)} The computing demands of \textit{pre-processing}, \textit{inference}, and \textit{post-processing} phases are decisive in choosing the optimal execution platform and machine type for it; \textbf{(B)} The parallelization techniques used in the programming of \textit{pre-processing} and \textit{post-processing} phases can directly impact the efficacy of scale-out; \textbf{(C)} The input data type of DL applications can amplify the overhead of virtualization; \textbf{(D)} Lower values of the container cores to the host cores (a.k.a. CHR) ratio remarkably increase the imposed overhead of vanilla CN; \textbf{(E)} Pinning CPU cores mitigates the imposed overhead of execution platforms, particularly, when the DL application is deployed on vanilla CN. Based on these findings, in the future, we plan to mathematically model the overhead of a given execution platform based on the application characteristics.

We also provide the following \textbf{Best Practices} to help solution architects efficiently deploy DL applications on Cloud:

\boxfig{
\noindent\textbf{(1)} Scrutinize resource utilization patterns of all three phases in the DL application before choosing an execution platform for deployment.

\noindent\textbf{(2)} Investigate the impact of scale-up on the E2E inference time for your specific DL application to find the optimal number of CPU cores. A 16-core instance is generally a good point to start this investigation.

\noindent\textbf{(3)} Do not scale-up to all CPU cores of a host. A few unused cores can mitigate the virtualization overhead. 

\noindent\textbf{(4)} Utilizing high-performance storage (\eg SSD) improves the usability of DL applications with many small-size input files.

\noindent\textbf{(5)} If you would like to benchmark DL applications to figure out the optimal deployment, make sure it measures the E2E inference time and not only the time for the \textit{inference} operation.

\noindent\textbf{(6)} If pre- and post-processing phases appear to dominate the E2E inference time, utilizing GPU is not necessarily helpful, instead, increasing the scale-up level may offer a better cost/performance trade-off. 

\noindent\textbf{(7)} Container platform imposes the least overhead and is suggested for deploying DL applications on Edge.

}
\balance 
\bibliographystyle{./bibliography/IEEEtran}
\bibliography{./bibliography/IEEEabrv,./bibliography/IEEEexample}

\begin{thebibliography}{10}
\providecommand{\url}[1]{#1}
\csname url@samestyle\endcsname
\providecommand{\newblock}{\relax}
\providecommand{\bibinfo}[2]{#2}
\providecommand{\BIBentrySTDinterwordspacing}{\spaceskip=0pt\relax}
\providecommand{\BIBentryALTinterwordstretchfactor}{4}
\providecommand{\BIBentryALTinterwordspacing}{\spaceskip=\fontdimen2\font plus
\BIBentryALTinterwordstretchfactor\fontdimen3\font minus
  \fontdimen4\font\relax}
\providecommand{\BIBforeignlanguage}[2]{{%
\expandafter\ifx\csname l@#1\endcsname\relax
\typeout{** WARNING: IEEEtran.bst: No hyphenation pattern has been}%
\typeout{** loaded for the language `#1'. Using the pattern for}%
\typeout{** the default language instead.}%
\else
\language=\csname l@#1\endcsname
\fi
#2}}
\providecommand{\BIBdecl}{\relax}
\BIBdecl

\bibitem{salehi2014resource}
M.~A. Salehi, B.~Javadi, and R.~Buyya, ``Resource provisioning based on
  preempting virtual machines in distributed systems,'' \emph{Concurrency and
  Computation: Prac. and Exper.}, vol.~26, no.~2, pp. 412--433, 2014.

\bibitem{agache2020firecracker}
A.~Agache, M.~Brooker, A.~Iordache, A.~Liguori, R.~Neugebauer, P.~Piwonka, and
  D.-M. Popa, ``Firecracker: Lightweight virtualization for serverless
  applications,'' in \emph{17th USENIX symposium on networked systems design
  and implementation (NSDI 20)}, 2020, pp. 419--434.

\bibitem{pahl2015containerization}
C.~Pahl, ``{Containerization and the PaaS cloud},'' \emph{IEEE Cloud
  Computing}, vol.~2, no.~3, pp. 24--31, 2015.

\bibitem{samani2020art}
D.~G. Samani, C.~Denninnart, J.~Bacik, and M.~Amini~Salehi, ``The art of
  cpu-pinning: Evaluating and improving the performance of virtualization and
  containerization platforms,'' in \emph{Proceedings of the 49th International
  conference on parallel processing}, ser. ICPP '20, 2020.

\bibitem{lazaro2019approach}
O.~D.~M. L{\'a}zaro, W.~M. Mohammed, B.~R. Ferrer, R.~Bejarano, and J.~L.~M.
  Lastra, ``An approach for adapting a cobot workstation to human operator
  within a deep learning camera,'' in \emph{Proceedings of the 17th IEEE
  International Conference on Industrial Informatics}, ser. INDIN '19,
  vol.~1.\hskip 1em plus 0.5em minus 0.4em\relax IEEE, 2019, pp. 789--794.

\bibitem{devlin2018bert}
J.~Devlin, M.-W. Chang, K.~Lee, and K.~Toutanova, ``Bert: Pre-training of deep
  bidirectional transformers for language understanding,'' \emph{arXiv preprint
  arXiv:1810.04805}, 2018.

\bibitem{reddi2020mlperf}
V.~J. Reddi, C.~Cheng, D.~Kanter, P.~Mattson, G.~Schmuelling, C.-J. Wu,
  B.~Anderson, M.~Breughe, M.~Charlebois, and Chou, ``Mlperf inference
  benchmark,'' in \emph{Proceedings of the 47th ACM/IEEE Annual International
  Symposium on Computer Architecture (ISCA)}, 2020, pp. 446--459.

\bibitem{PinningHPCCLab}
\BIBentryALTinterwordspacing
D.~G.Samani, ``{CPU Pinning for KVM and Docker}.'' [Online]. Available:
  \url{https://tinyurl.com/a4rbb4}
\BIBentrySTDinterwordspacing

\bibitem{reddi2020mlperfVision}
MLPerf, ``Classification,''
  \url{https://github.com/mlcommons/inference/tree/master/vision/classification_and_detection},
  Accessed Oct '21.

\bibitem{reddi2020mlperfSpeech}
MLPerf., ``Speech recognition,''
  \url{https://github.com/mlcommons/inference/tree/master/speech_recognition/rnnt},
  Accessed Oct '21.

\bibitem{reddi2020mlperfLanguage}
MLPerf, ``Natural language processing,''
  \url{https://github.com/mlcommons/inference/tree/master/language/bert},
  Accessed Nov '21.

\bibitem{reddi2020mlperfRecommendation}
------, ``Recommendation, pytorch,''
  \url{https://github.com/mlcommons/inference/tree/master/recommendation/dlrm/pytorch},
  Accessed Oct '21.

\bibitem{denninnart2021harnessing}
C.~Denninnart and M.~A. Salehi, ``Harnessing the potential of function-reuse in
  multimedia cloud systems,'' \emph{IEEE Transactions on Parallel and
  Distributed Systems (TPDS), Volume 33, Issue 3, Pages: 617– 629}, 2021.

\bibitem{madhavapeddy2013unikernels}
A.~Madhavapeddy, R.~Mortier, C.~Rotsos, D.~Scott, B.~Singh, T.~Gazagnaire,
  S.~Smith, S.~Hand, and J.~Crowcroft, ``Unikernels: Library operating systems
  for the cloud,'' \emph{ACM SIGARCH Computer Architecture News}, vol.~41,
  no.~1, pp. 461--472, 2013.

\bibitem{leon2020dark}
M.~Leon, ``The dark side of unikernels for machine learning,'' \emph{arXiv
  preprint arXiv:2004.13081}, 2020.

\bibitem{potdar2020performance}
A.~M. Potdar, D.~Narayan, S.~Kengond, and M.~M. Mulla, ``Performance evaluation
  of docker container and virtual machine,'' \emph{Procedia Computer Science},
  vol. 171, pp. 1419--1428, 2020.

\bibitem{lingayat2018performance}
A.~Lingayat, R.~R. Badre, and A.~K. Gupta, ``Performance evaluation for
  deploying docker containers on baremetal and virtual machine,'' in
  \emph{Proceedings of the 3rd International Conference on Communication and
  Electronics Systems}, ser. ICCES '3.\hskip 1em plus 0.5em minus 0.4em\relax
  IEEE, 2018, pp. 1019--1023.

\bibitem{devarajan2021dlio}
H.~Devarajan, H.~Zheng, A.~Kougkas, X.-H. Sun, and V.~Vishwanath, ``Dlio: A
  data-centric benchmark for scientific deep learning applications,'' in
  \emph{Proceedins of the 21st IEEE/ACM International Symposium on Cluster,
  Cloud and Internet Computing}, ser. CCGrid '21.\hskip 1em plus 0.5em minus
  0.4em\relax IEEE, 2021, pp. 81--91.

\bibitem{koirala2019deep}
A.~Koirala, K.~Walsh, Z.~Wang, and C.~McCarthy, ``Deep learning for real-time
  fruit detection and orchard fruit load estimation: Benchmarking of
  ‘mangoyolo’,'' \emph{Precision Agriculture}, vol.~20, no.~6, pp. 45--60,
  2019.

\bibitem{duan2016benchmarking}
Y.~Duan, X.~Chen, R.~Houthooft, J.~Schulman, and P.~Abbeel, ``Benchmarking deep
  reinforcement learning for continuous control,'' in \emph{proceedings of the
  International conference on machine learning}.\hskip 1em plus 0.5em minus
  0.4em\relax PMLR, 2016, pp. 1329--1338.

\bibitem{mase2020benchmarking}
J.~M. Mase, P.~Chapman, G.~P. Figueredo, and M.~T. Torres, ``Benchmarking deep
  learning models for driver distraction detection,'' in \emph{proceedings of
  the International Conference on Machine Learning, Optimization, and Data
  Science}.\hskip 1em plus 0.5em minus 0.4em\relax Springer, 2020, pp.
  103--117.

\bibitem{zobaed2021saed}
S.~M. Zobaed, M.~A. Salehi, and R.~Buyya, ``{SAED: Edge-Based Intelligence for
  Privacy-Preserving Enterprise Search on the Cloud},'' ser. CCGrid '21, 2021.

\bibitem{fagbohungbe2021benchmarking}
O.~Fagbohungbe and L.~Qian, ``Benchmarking inference performance of deep
  learning models on analog devices,'' in \emph{Proceedings of the
  International Joint Conference on Neural Networks}, 2021, pp. 1--9.

\bibitem{liu2018benchmarking}
L.~Liu, Y.~Wu, W.~Wei, W.~Cao, S.~Sahin, and Q.~Zhang, ``Benchmarking deep
  learning frameworks: Design considerations, metrics and beyond,'' in
  \emph{Proceedings of the International Conference on Distributed Computing
  Systems}, ser. ICDCS '38.\hskip 1em plus 0.5em minus 0.4em\relax IEEE, 2018,
  pp. 1258--1269.

\bibitem{dai2019benchmarking}
W.~Dai and D.~Berleant, ``Benchmarking contemporary deep learning hardware and
  frameworks: A survey of qualitative metrics,'' in \emph{Proceedings of the
  International Conference on Cognitive Machine Intelligence}, ser. CogMI
  '19.\hskip 1em plus 0.5em minus 0.4em\relax IEEE, 2019, pp. 148--155.

\bibitem{wang2019benchmarking}
Y.~E. Wang, G.-Y. Wei, and D.~Brooks, ``Benchmarking tpu, gpu, and cpu
  platforms for deep learning,'' \emph{arXiv preprint arXiv:1907.10701}, 2019.

\bibitem{lin2018comparison}
C.-Y. Lin, H.-Y. Pai, and J.~Chou, ``Comparison between bare-metal, container
  and vm using tensorflow image classification benchmarks for deep learning
  cloud platform.'' in \emph{Proceedings of the International Conference on
  Cloud Computing and Services Science}, ser. CLOSER, 2018, pp. 376--383.

\bibitem{reddi2019mlperf}
T.~M. Group, ``Mlperf inference benchmark,'' 2019.

\bibitem{dlbsGithub}
DLBS, ``Deep learning benchmarking suite,''
  \url{https://github.com/HewlettPackard/dlcookbook-dlbs}, Accessed on 2021 Oct
  12.

\bibitem{deepbenchGithub}
DeepBench, ``Benchmarking deep learning on different hardware,''
  \url{https://github.com/baidu-research/DeepBench}, Accessed May '21.

\bibitem{yang2012using}
C.-T. Yang, H.-Y. Wang, and Y.-T. Liu, ``{Using PCI pass-through for GPU
  virtualization with Cuda},'' in \emph{Proceedings of the IFIP International
  Conference on Network and Parallel Computing}.\hskip 1em plus 0.5em minus
  0.4em\relax Springer, 2012, pp. 445--452.

\bibitem{liu2010understanding}
F.~Liu and Y.~Solihin, ``Understanding the behavior and implications of context
  switch misses,'' \emph{ACM Transactions on Architecture and Code Optimization
  (TACO)}, vol.~7, no.~4, pp. 1--28, 2010.

\bibitem{majo2011memory}
Z.~Majo and T.~R. Gross, ``{Memory management in NUMA multicore systems:
  trapped between cache contention and interconnect overhead},'' in
  \emph{Proceedings of the international symposium on Memory management}, ser.
  ISMM '11, 2011, pp. 11--20.

\bibitem{wong2008fairness}
C.~Wong, I.~Tan, R.~Kumari, J.~Lam, and W.~Fun, ``Fairness and interactive
  performance of o(1) and cfs linux kernel schedulers,'' in \emph{Proceedings
  of the International Symposium on Information Technology}, ser. ISMM '4, Aug.
  2008, pp. 1--8.

\bibitem{Josef}
J.~Bacik, ``{IO} and cgroups, the current and future work.''\hskip 1em plus
  0.5em minus 0.4em\relax Boston, MA: {USENIX} Association, feb 2019.

\bibitem{shirinbab2017performance}
S.~Shirinbab, L.~Lundberg, and E.~Casalicchio, ``Performance evaluation of
  container and virtual machine running cassandra workload,'' in
  \emph{Proceedings of the 3rd International Conference of Cloud Computing
  Technologies and Applications}, ser. CloudTech '17, Oct. 2017, pp. 1--8.

\end{thebibliography}

\end{document}